\documentclass[twocolumn,prl,preprintnumbers,superscriptaddress,amsmath,amssymb,english]{revtex4}
\usepackage{amsmath,amssymb,amsfonts,mathrsfs}
\usepackage{amsthm}
\usepackage{CJK}
\usepackage{bm}
\usepackage{babel}
\usepackage{graphicx}
\allowdisplaybreaks
\usepackage{braket}
\usepackage[breaklinks]{hyperref}
\usepackage{color}
\usepackage{epstopdf}
\hypersetup{colorlinks=true, linkcolor=blue, citecolor=blue, filecolor=blue, urlcolor=blue}

\begin{document}

\title{Flexible Control of Chiral Superconductivity in Optically Driven Nodal Point Superconductors with Antiferromagnetism}


\author{Zhen Ning}
\affiliation{Institute for Structure and Function $\&$ Department of Physics $\&$ Chongqing Key Laboratory for Strongly Coupled Physics, Chongqing University, Chongqing 400044, People's Republic of China}
\affiliation{Center of Quantum Materials and Devices, Chongqing University, Chongqing 400044, People's Republic of China}

\author{Junjie Zeng}
\affiliation{Institute for Structure and Function $\&$ Department of Physics $\&$ Chongqing Key Laboratory for Strongly Coupled Physics, Chongqing University, Chongqing 400044, People's Republic of China}
\affiliation{Center of Quantum Materials and Devices, Chongqing University, Chongqing 400044, People's Republic of China}

\author{Da-Shuai Ma}
\affiliation{Institute for Structure and Function $\&$ Department of Physics $\&$ Chongqing Key Laboratory for Strongly Coupled Physics, Chongqing University, Chongqing 400044, People's Republic of China}
\affiliation{Center of Quantum Materials and Devices, Chongqing University, Chongqing 400044, People's Republic of China}

\author{Dong-Hui Xu}
\email[]{donghuixu@cqu.edu.cn}
\affiliation{Institute for Structure and Function $\&$ Department of Physics $\&$ Chongqing Key Laboratory for Strongly Coupled Physics, Chongqing University, Chongqing 400044, People's Republic of China}
\affiliation{Center of Quantum Materials and Devices, Chongqing University, Chongqing 400044, People's Republic of China}

\author{Rui Wang}
\email[]{rcwang@cqu.edu.cn}
\affiliation{Institute for Structure and Function $\&$ Department of Physics $\&$ Chongqing Key Laboratory for Strongly Coupled Physics, Chongqing University, Chongqing 400044, People's Republic of China}
\affiliation{Center of Quantum Materials and Devices, Chongqing University, Chongqing 400044, People's Republic of China}

\date{\today}

\begin{abstract}
Recent studies have attracted widespread attention on magnet-superconductor hybrid systems with emergent topological superconductivity.
Here, we present the Floquet engineering of realistic two-dimensional topological nodal-point superconductors that are composed of antiferromagnetic monolayers in proximity to an $s$-wave superconductor.
We show that Floquet chiral topological superconductivity arises naturally due to light-induced breaking of the effective time-reversal symmetry. More strikingly, we find that the Floquet chiral topological superconducting phases can be flexibly controlled by irradiating elliptically polarized light, with the photon-dressed quasienergy spectrum carrying different Chern numbers. Such optically switchable topological transition is attributed to the simultaneous creations (or annihilations) of valley pairs. 
Our findings provide a feasible approach for achieving the Floquet chiral topological superconductivity with flexible tunability, which would draw extensive attention in experiments.
\end{abstract}
\maketitle

\emph{{\color{magenta}Introduction.}}---The exploration of topological phases of matter is one of extensive activities in the modern condensed-matter community~\cite{TI1,TI2}. Among various topological phases, topological superconductors (TSCs), which host topologically protected Majorana quasiparticles (or termed as Majorana modes), are the subject of intense recent studies~\cite{MF1,MF2,MF3,MF4,MF5,MF6}. In the TSC systems, Majorana zero modes are manifested as zero-energy end-localized bound states in one-dimensional (1D) cases or vortex bound states in two-dimensional (2D) cases. Importantly, Majorana zero modes obey non-Abelian statistics and are proposed as a fundamental building blocks for topological quantum information processing~\cite{Ivanov2001,tqc1,tqc2,tqc3,tqc4,tqc5}, bridging the gap between condensed matter physics and quantum computations. Hence, creation and manipulation of Majorana modes at experimentally feasible levels are of particular importance. Similar to topological phases in normal states, the investigation of topological superconductivity has not only been limited to gapped TSCs but also extended to gapless topological nodal superconductors (TNSCs) topologically protected gapless quasiparticle excitations in the bulk as well as distinctive gapless Majorana modes at the surface or edge~\cite{NSC1,NSC2,NSC3,NSC4,NSC5,NSC6,NSC7,NSC8,NSC9,NSC10,NSC11,NSC12,NSC13,NSC14,NSC15,NSC16,RevModPhys2016}.


It is worth noting that Floquet engineering with periodic diving offers a powerful tool for engineering quantum states with high tunability due its dynamic manipulation on ultrafast time scales \cite{Shirley,Goldman,Eckardt2015,Bukov2015,Oka2019}. The periodic driving can not only potentially generate nontrivial band topology but also other exotic quantum phenomena that are absent in the static systems \cite{exp1,exp2,exp3,Rev1,Rev2,Rev3}. In the past decade, Floquet engineering has been extensively employed to study various quantum systems exhibiting intriguing topological features \cite{FT1,FT2,FSM1,FSM2,FSM3,FSM4,FSM5}. Remarkably, periodic-driven TSCs can enable anomalous Floquet Majorana modes beyond the Floquet topological band theory \cite{AES,FMaj1,FMaj2,FMaj3,FMaj4,FMaj5,FMaj6,FMaj7,FMaj8,FMaj9,FMaj10}, and thus lots of exotic phenomena related to Floquet TSCs have inspired tremendous research activity \cite{FTSC1,FTSC2,FTSC3,FTSC4,FTSC5,FTSC6}. While recent theoretical advancements have been very encouraging, the experimental exploration of Floquet TSCs is very slow since realistic material systems with topological superconductivity are rather scarce. Moreover, so far, most attempts have concentrated towards engineering Floquet Majorana modes in gapped TSCs, but the investigation of periodically driven TNSCs has been largely unexplored. Especially, the TNSCs host symmetry-protected nodal points or nodal lines, and thus light control of symmetry utilizing photon-dressed quasienergy spectrum with pairing correlations is expected to give rise to more abundant Floquet Majorana physics.

\begin{figure}[htb]
	\centering
	\includegraphics[width=0.45\textwidth]{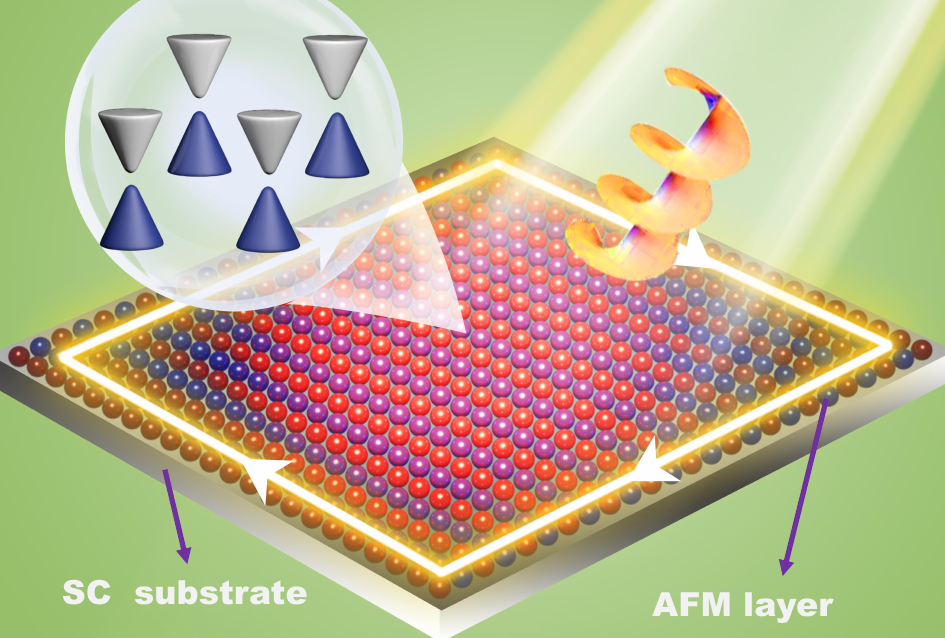}
	\caption{The schematic illustration of the TNPSC system under light irradiation. The antiferromagnetic monolayer is placed on the superconductor substrate. The sublattices in each unite cell are separated by different colors (spin up $\uparrow$ (red), spin down $\downarrow$ (blue)). The incident light is elliptically polarized with $\bm{A}(t) = [A_x\cos(\omega t),A_y\sin(\omega t)]$ which transform the nodal-point system into a gapped FCTSC. The chiral edge state is highlighted on the boundary of the system.}
	\label{Fig:Fig1}
\end{figure}

\begin{figure}[htb]
	\centering
	\includegraphics[width=0.5\textwidth]{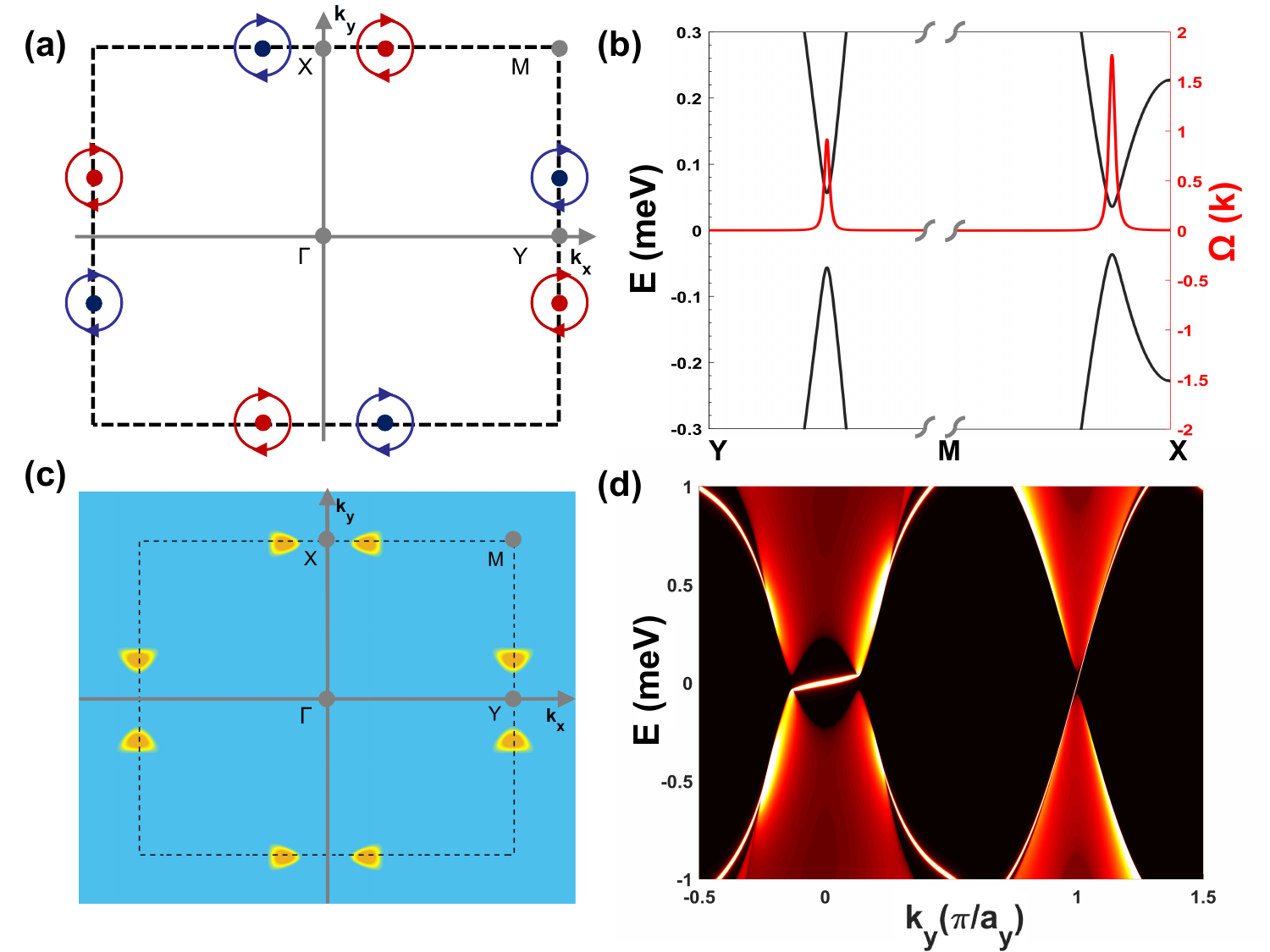}%
	\caption{ (a) The nodal points of TNPSC phase at the boundaries of magnetic Brillouin zone in the absence of light irradiation. The chirality of each nodal point is denoted by circles with different colors ($+1$ red, $-1$ blue). 
The high symmetry points are also denoted. (b) The quasiparticle band structures and Berry curvature along the high-symmetry $Y$-$M$-$X$. (c) The distribution of $\Omega(\mathbf{k})$ in whole Brillouin zone, indicating that the nonzero $\Omega(\mathbf{k})$ diverges at the valleys. 
(d) The local density of states of a semi-infinite nanoribbon with the periodic boundary in the $k_y$  direction. In panels (b)-(d), we set the polar angle to $\theta = \frac{\pi}{4}$ (i.e., CPL) and light intensity $k_A = 0.8$. }
	\label{Fig:Fig2}
\end{figure}

Here, we present Floquet engineering of realistic topological nodal-point superconductors (TNPSCs) with antiferromagnetism under irradiation of elliptically polarized light (see Fig. \ref{Fig:Fig1}), and Floquet chiral topological superconductor (FCTSCs) that can be experimentally accessed is achieved. Such TNPSCs have been realized in a magnet-superconductor hybrid system, i.e., antiferromagnetic monolayers on top of an $s$-wave superconductor \cite{TNPSC}. Recently, the magnet-superconductor hybrid systems have been expected to provide a feasible avenue for the experimental identification of Majorana zero modes and thus have attracted intense attention~\cite{MHS1,MHS2,MHS3,MHS4,MHS5,MHS6,MHS7,MHS8}, due to the interplay of magnetic order and unconventional pairing correlations. 
Without light irradiation, the antiferromagnetic TNPSC hosts gapless symmetry-protected nodal points that are featured by specific chirality and located at the boundary of magnetic Brillouin zone [see Fig. \ref{Fig:Fig2}(a)]. In the presence of light irradiation,
light-induced symmetry breaking can generally gap out the nodal points  (see Fig. \ref{Fig:Fig1}). 
Due to the magnetic and crystal anisotropy, the gapped nodal points form two pairs of non-equivalent valleys. By manipulating the intensity and polar angle of elliptically polarized light, we can selectively and simultaneously create and annihilate a pair of valleys. The corresponding merging transition of valley pairs accompanies with gap closing and reopening, driving the system towards FCTSC phases from gapless TNPSC states. Remarkably, the FCTSC phases can be flexibly controlled, with the photon-dressed quasienergy spectrum carrying different Chern numbers.
Focusing on the experimentally achieved TNPSC, light control of FCTSCs intertwining with antiferromagnetic order, superconductivity, and topology would draw extensive attention.

\emph{{\color{magenta}Model and theory.}}---Our starting point is the 2D TNPSC with antiferromagnetism, i.e., antiferromagnetic monolayers crystallizing in a
 rectangular lattice on top of an $s$-wave superconductor (see Fig. \ref{Fig:Fig1}), and the tight-binding Hamiltonian is given by~\cite{TNPSC, MHS1}
\begin{equation}
\begin{aligned}
H=&\sum_{\bm{r}}[\sum_{i=0,1,2}t_ic^{\dag}_{\bm{r}+\bm{\delta}_i}c_{\bm{r} }
-\sum_{i=0,1}i\alpha c^{\dag}_{\bm{r}+\bm{\delta}_i}(\bm{\sigma}\times \bm{e}_i)_zc_{\bm{r}}\\
+&Je^{i\bm{Q}\cdot\bm{r}}c^{\dag}_{\bm{r} }\sigma_zc_{\bm{r}}
+\Delta(c^{\dag}_{\bm{r} \uparrow}c^{\dag}_{\bm{r}\downarrow}
+ c_{\bm{r} \uparrow}c_{\bm{r}\downarrow})-\mu  c^{\dag}_{\bm{r}}c_{\bm{r}}],\\
\end{aligned}
\label{eq:TBH-0}
\end{equation}
where $c^{\dag}_{\bm{r}} = [c^{\dag}_{\bm{r}\uparrow},c^{\dag}_{\bm{r}\downarrow}]$ and $c_{\bm{r}}$ are two component creation and annihilation operators with spin degrees of freedom ($\uparrow,\downarrow$) at the site $\mathbf{r}$, respectively. 
As shown in Fig. \ref{Fig:Fig1}, the unit cell consists of two sublattices, each hosting magnetic atoms with antiparallel spins. We set the lattice vectors $\bm{a}_x = (a,0)$,  $\bm{a}_y = (0,b)$ with two lattice constants $a$ and $b$.
The nearest, next-nearest, and next-next-nearest neighbor vectors are expressed as $\bm{\delta}_0 = (\frac{a}{2},\frac{b}{2})$, $\bm{\delta}_1 = (a,0)$, and $\bm{\delta}_2 = (0,b)$, and their corresponding hopping amplitudes are respectively denoted by $t_0$, $t_1$, and $t_2$.
The second term in Eq.~(\ref{eq:TBH-0}) is the Rashba spin-orbital coupling, where the parameter $\alpha$ denotes its strength and $\bm{e}_i = \bm{\delta}_i/|\bm{\delta}_i|$ is the unit vector of the nearest or next-nearest neighbor.
The antiferromagnetic exchange coupling, superconducting order parameter, and chemical potential are denoted by $J$, $\Delta$, and $\mu$, respectively. We set the values of all parameters in Eq.~(\ref{eq:TBH-0}) to be the same as in \rm{Ref.~\cite{TNPSC}}, and the topological nodal-point superconducting phase is reproduced as shown in Fig. \ref{Fig:Fig2}(a). The nodal-points carrying specific chirality are related to an effective time-reversal symmetry $\mathcal{S}=\mathcal{T}\tau_{1/2}$, which combines the time-reversal symmetry $\mathcal{T}$ with a half-translation operator $\tau_{1/2}=e^{i\mathbf{k}\cdot(\mathbf{a}_x+\mathbf{a}_y)/2}$. 
Through carrying the symmetry analysis of magnetic space group, we find that the chiral nodal-points are guaranteed by nonsymmorphic symmetry operators $\tilde{M}_i$ with two-fold rotation operators $\tilde{M}_i=\mathcal{S}C_{2i}$ ($i=x, y, z$); here, $\tilde{M}_z$ allows the nodal points rather nodal lines in the $k_x$-$k_y$ plane while $\tilde{M}_x$ and $\tilde{M}_y$ force these nodal points symmetrically distributed at the boundary of magnetic Brillouin zone.


Then, we consider the system subject to a space-homogeneous periodic-driving field characterized by the vector potential $\bm{A}(t) = [A_x\cos(\omega t),A_y\sin(\omega t)]$, where the frequency $\omega$ is related to one period $T = 2\pi/\omega$. In the following discussion, the elliptically polarized light is described by its amplitude $k_A = eA_0 = e\sqrt{A^2_x+A^2_y}$ and the angle $\theta = \arctan(A_y/A_x)$. The coupling between the system and light filed $\bm{A}(t)$ is introduced through the Peierls substitution, where the hopping amplitudes are attached to an additional phase factor $e^{ie\bm{A} \cdot (\bm{r}_i-\bm{r}_j)}$. To reveal this, we first rewrite the tight-binding Hamiltonian Eq. (\ref{eq:TBH-0}) into the standard form as
\begin{equation}
\begin{aligned}
H&=\sum_{\bm{R}}[\psi^{\dag}_{\bm{R}} T_{0}\psi_{\bm{R}}
+\sum_{i=1,2,3}\psi^{\dag}_{\bm{R}+\bm{\delta}_i} T_{i}\psi_{\bm{R}}+h.c.].\\
\end{aligned}
\label{eq:H0}
\end{equation}
Then, the photon-dressed hopping matrices $T_{0}$ and $T_{i}$ in the presence of light field can be explicitly written as
\begin{equation}
\begin{aligned}
T_0&\rightarrow\tau_z\eta_{+}\Lambda_{++}e^{i\phi_{++}(t)}
+ \tau_z\eta_{-}\Lambda_{--}e^{i\phi_{--}(t)}\\
&-\mu\tau_z\eta_0\sigma_0+J\tau_0\eta_z\sigma_z+\Delta\tau_x\eta_0\sigma_0\\
T_1&\rightarrow\tau_z\eta_{+}\Lambda_{-+}e^{i\phi_{-+}(t)}
+ \tau_z\eta_{0}(\sigma_0t_1-i\alpha\sigma_y)e^{-i\phi_{1}(t)}\\
T_2&\rightarrow \tau_z\eta_{+}\Lambda_{+-}e^{i\phi_{+-}(t)}
+ \tau_z\eta_{0}\sigma_0t_2e^{-i\phi_{2}(t)}\\
T_3&\rightarrow\tau_z\eta_{+}\Lambda_{--}e^{i\phi_{--}(t)},
\end{aligned}
\label{eq:T0123}
\end{equation}
where we introduce $\Lambda_{ss'} = \sigma_0t_0-i\alpha(\boldsymbol{\sigma}\times \boldsymbol{\delta}_{ss'})_z$ with $\boldsymbol{\delta}_{ss'} = (s\frac{a_x}{2},s'\frac{a_y}{2})$, $s = \pm$. The time-dependent phase factors are $\phi_{ss'}(t) = e\mathbf{A}(t)\cdot\boldsymbol{\delta}_{ss'}$ and $\phi_{1,2}(t) = e\mathbf{A}(t)\cdot\boldsymbol{\delta}_{1,2}$.

The periodically driven Hamiltonian Eqs. (\ref{eq:H0}) and (\ref{eq:T0123}) can be solved by the Floquet theorem \cite{Goldman,Oka2019}; 
that is, we transform the time-dependent Hamiltonian $H(t) = H(t + T)$ into frequency domain and obtain the time-independent Floquet Hamiltonian as $H^F_{nm} = n\omega\delta_{nm}  + h^F_{n-m}$, where the Fourier components are $h^F_{mn}=\frac{1}{T}\int^T_0 dt H(t) e^{i\omega (m-n)}$. Then we obtain the Floquet-Bloch states $|u^n(\mathbf{k})\rangle$ and the quasienergies $\epsilon_n(\mathbf{k})$ by diagonalizing the Floquet Hamiltonian,
\begin{equation}
\begin{aligned}
H^F|u^n_{\mathbf{k}}\rangle = \epsilon_n(\mathbf{k})|u^n_{\mathbf{k}}\rangle.
\end{aligned}
\label{eq:HF}
\end{equation}
In this work, we focus on the low energy regime near the nodal point, the typical energy scale is $\epsilon_0 \sim 1 meV$. We consider the off-resonant regime and choose a high frequency of the light field $\hbar \omega = 20meV\gg\epsilon_0$, and thus the process of band transitions induced by directly absorbing photons can be ignored. 


\emph{{\color{magenta}Light induced Floquet chiral topological superconductor.}}---
In the presence of light field, the quasiparticle spectra are effectively renormalized  by the virtual photon absorption processes. To be specific, we fist set the polar angle to $\theta = \frac{\pi}{4}$ and this case corresponds to circularly polarized light (CPL). The quasienergies $\epsilon_n(\mathbf{k})$ and the eigenstates $|u^n_{\mathbf{k}}\rangle$ are obtained from the Eq.~(\ref{eq:HF}). Under the irradiation of CPL, the effective time-reversal symmetry $\mathcal{S}$ is explicitly broken, and then the absence of $\tilde{M}_z$ will gap out the nodal points. As shown in Fig. \ref{Fig:Fig2}(b), we plot the quasiparticle band structures with the amplitude of light field $k_A = 0.8$ along the high-symmetry $Y$-$M$-$X$. Because of the magnetic and crystal anisotropy, the bands show different gaps along the $M$-$X$ and $M$-$Y$ directions, forming two non-equivalent valleys. In order to characterize the topological feature of light-induced gapped phase,
we introduce the Berry curvature of quasiparticle spectra in momentum space
\begin{equation}
\begin{aligned}
\Omega_n(\mathbf{k}) = -2\sum_{m\ne n}\mathrm{Im}
\frac{\langle u^n_{\mathbf{k}}|\partial_{k_x}H^F|u^m_{\mathbf{k}}\rangle\langle u^m_{\mathbf{k}}|\partial_{k_y}H^F|u^n_{\mathbf{k}}\rangle}
{[\epsilon_n(\mathbf{k})-\epsilon_m(\mathbf{k})]^2},
\end{aligned}
\label{eq:BC}
\end{equation}
The band topology can be characterized by the summation of Berry curvature from all the occupied quasiparticle bands $\Omega(\mathbf{k}) = \sum_{n\in \mathrm{occ}}\Omega_n(\mathbf{k})$.
In Fig. \ref{Fig:Fig2}(b), we depict the corresponding Berry curvature illustrated as the red curves, which show peaks around the valleys. We also plot the distribution of $\Omega(\mathbf{k})$ in whole Brillouin zone [see Fig. \ref{Fig:Fig2}(c)]. We can see that the nonzero $\Omega(\mathbf{k})$ diverges at all valleys, indicating the topologically non-triviality of these gaps opened by light irradiation. To verify the topological category this light-induced gapped superconducting phase, we calculate the Chern number from Eq.~(\ref{eq:BC}) as
\begin{equation}
\begin{aligned}
\mathcal{N} = \sum_{n\in \mathrm{occ}}\int_{BZ,\mathbf{k}}\Omega_n(\mathbf{k}).
\end{aligned}
\label{eq:ChernN}
\end{equation}
By summing the contributions of $\Omega(\mathbf{k})$ from the occupied bands, we find that each valley contributes a half-integer Chern number. Thus, the system become a FCTSC with Chern number $\mathcal{N} = 2$.


According to the bulk-boundary correspondence, the topological feature of the FCTSC can also be depicted by chiral Majorana edge modes.
Therefore, we further explore these phenomena by calculating the local density of states using the iterative Green's function method \cite{RGFM}. We consider a semi-infinite nanoribbon imposing the open boundary in the $x$ direction and the periodic boundary in the $k_y$  direction.
As shown in Fig. \ref{Fig:Fig2}(d), as expected, there are two chiral Majorana edge modes in the bulk gap, coinciding with the FCTSC with $\mathcal{N} = 2$.

\begin{figure}[htb]
	\centering
	\includegraphics[width=0.49\textwidth]{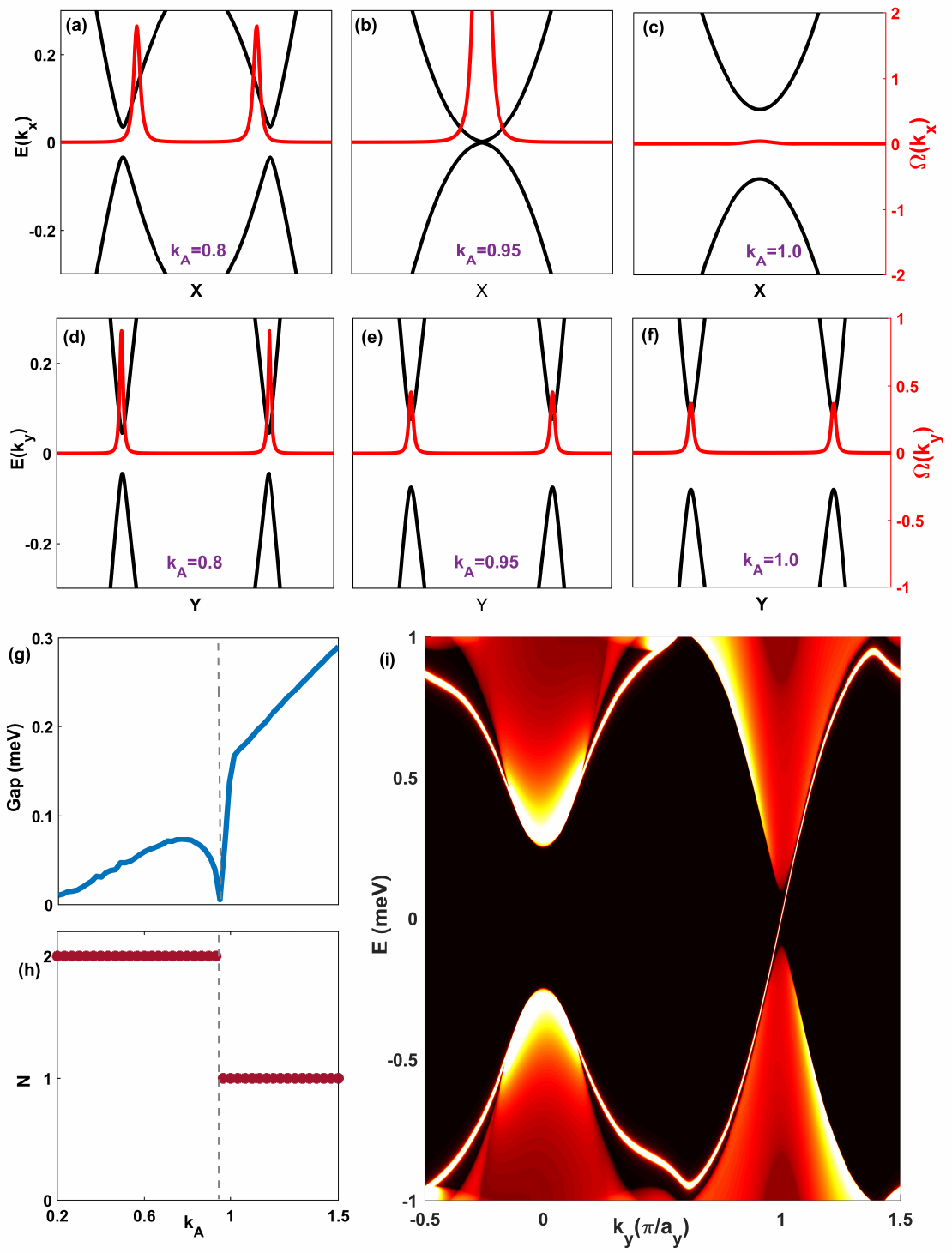}%
	\caption{Light-induced evolution of Floquet quasiparticle bands and their corresponding Berry curvature along the boundary of (a)-(c) $k_y=\frac{\pi}{b}$ or (d)-(f) $k_x=\frac{\pi}{a}$. (g) The variation of gaps and (h) Chern number with the increase of light intensity. (i) The local density of states of a semi-infinite nanoribbon with the periodic boundary in the $k_y$  direction.
}
	\label{Fig:Fig3}
\end{figure}


\emph{{\color{magenta} Flexible light control of Floquet chiral superconductivity.
}}---As discussed above, the light-irradiated TNPSC states can be transformed into a FCTSC phase with $\mathcal{N} = 2$ under a periodic driving of CPL. 
Moreover, as shown in Fig. \ref{Fig:Fig2}(b), it is worth noting that the irradiation of light field induces explicit valley polarization (i.e., there are different gaps along $M$-$X$ and $M$-$Y$) since the light-induced symmetry breaking is accompanied by magnetic and crystal anisotropy. Therefore, the non-equivalent valleys are expected to possess different evolution behaviours by manipulating the intensity $k_A$ and polar angle $\theta$ of incident light $\mathbf{A}(t)$. To illustrate this, we first examine the light-induced evolution of Floquet quasiparticle bands and their corresponding $\Omega(\mathbf{k})$ along the boundary of magnetic Brillouin zone at some typical values of light intensity with a fixed polar angle of $\theta=\frac{\pi}{4}$, as shown in Figs. \ref{Fig:Fig3}(d)-\ref{Fig:Fig3}(f). With the increase of light intensity $k_A$, the pair of valleys at the boundary with $k_y=\frac{\pi}{b}$ move towards to each other and merge at the $X$ point [see Figs. \ref{Fig:Fig3}(a)-\ref{Fig:Fig3}(c)]; while those at the boundary with $k_x=\frac{\pi}{a}$ are nearly maintained  [see Figs. \ref{Fig:Fig3}(d)-\ref{Fig:Fig3}(f)]. The gaps close and reopen when the light intensity increases beyond the critical value $k_A=0.95$ [see Figs. \ref{Fig:Fig3}(b) and \ref{Fig:Fig3}(g)]. The merging process of one pair of valleys at the boundary with $k_y=\frac{\pi}{b}$ is accompanied by a topological transition, and the Chern number is reduced to $\mathcal{N}  = 1$ [see Fig. \ref{Fig:Fig3}(h)]. We further calculate the local density of states as shown in Fig. \ref{Fig:Fig3}(i), which present only one chiral Majorana edge mode inside the bulk gap.

\begin{figure}[htb]
	\centering
	\includegraphics[width=0.49\textwidth]{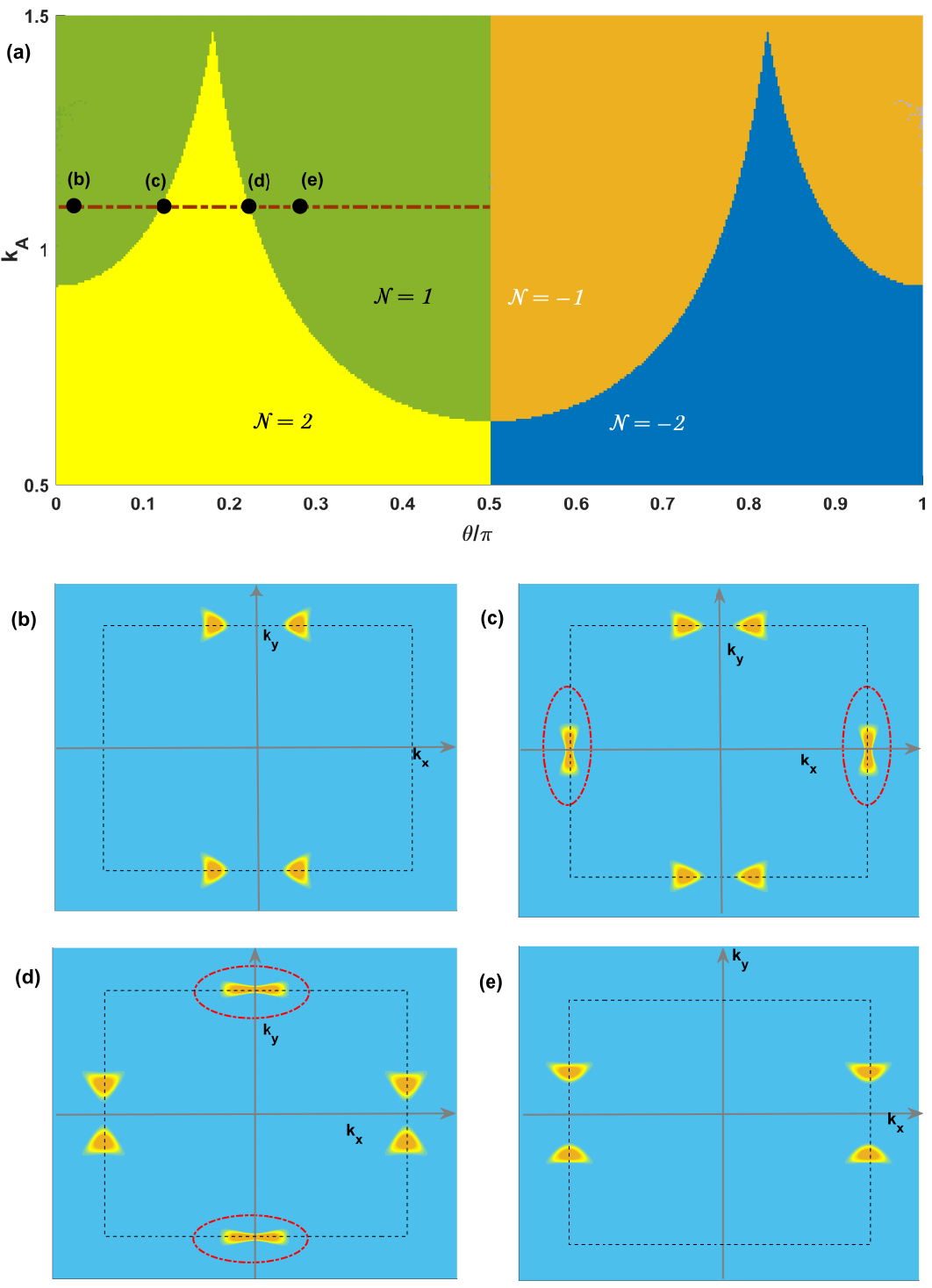}%
	\caption{(a) The topological phase diagram  depicted by the Chern numbers on the $\theta$-$k_A$ parameter plane. 
In panels (b)-(e), the distributions of the Berry curvature $\Omega(\mathbf{k})$ in momentum space. We fix the light intensity at $k_A = 1.1$ and choose the polar angle $\theta/\pi = 0.01$, $0.13$, $ 0.21$, and $ 0.3$ respectively, see the dashed line and the solid points in panel (a).
}
	\label{Fig:Fig4}
\end{figure}

Thus far, we have only studied the case of a system subjected to CPL (i.e., fixing $\theta=\frac{\pi}{4}$). Considering the magnetic and crystal anisotropy, a further study on the coherent interactions between the light intensity and polar angle $\theta$ is desirable.
To fully characterize the light-induced topological superconducting phases, we present a comprehensive phase diagram depicted by the Chern numbers on
the $\theta$-$k_A$ parameter plane, as shown in Fig. \ref{Fig:Fig4}(a). It is found that the critical value of light intensity near the phase transition is dependent on the polar angle $\theta$. The topological transition can be easily induced when $\theta$ is close to $\frac{\pi}{2}$. Otherwise, the induction of topological transition requires a higher value of light intensity. The interval $\theta\in (0,\frac{\pi}{2})$ and $\theta\in (\frac{\pi}{2},\pi)$ are identical except that the signs of the Chern numbers are opposite, indicating that the chiral Majorana edge modes can be reversed by changing the handedness of incident elliptically polarized light. According to the phase diagram, one can achieve a series of topological phase transitions by varying the intensity and polar angle. More strikingly, the pair of valleys on the different boundaries can be selectively controlled. 
We use the distribution of $\Omega(\mathbf{k})$ to present an overall look of the movement of the valleys at the boundaries of Brillouin zone. In Figs. \ref{Fig:Fig4}(b)-\ref{Fig:Fig4}(e), we fix the light intensity at $k_A = 1.1$ and check the evolution of the valleys with different values of $\theta$. It is found that the first transition occurs at a critical angle $\theta_c\approx 0.13\pi$, beyond which a pair of valleys at the boundary of $k_x=\frac{\pi}{a}$ are created [see Fig. \ref{Fig:Fig4} (c)], accompanying with Chern number jumping from $\mathcal{N} = 1\rightarrow 2$. By further increasing the polar angle $\theta$, the pair of valleys at the boundary of $k_y=\frac{\pi}{b}$ move close to each other and will annihilate at a critical point $\theta_c\approx 0.21\pi$ [see Fig. \ref{Fig:Fig4}(d)], thereby leaving only one pair of valleys in the Brillouin zone as shown in Fig. \ref{Fig:Fig4}(e). Therefore, we conclude that the FCTSC phases with the Chern numbers ranging from $\mathcal{N}=-2$ to $\mathcal{N}=2$ can be flexibly controlled, which is attributed to the simultaneous creations (or annihilations) of valley pairs by manipulating the intensity and polar angle of incident light.



\emph{{\color{magenta}Summary.}}--- In summary, we have demonstrated that the FCTSC phases with flexible tunability can be achieved in the periodically driven magnet-superconductor hybrid systems, which initially possess TNPSC states with antiferromagnetism. Under irradiation of elliptically polarized light, we show that manipulation of intensity and polar angle can selectively and simultaneously create (or annihilate) valley pairs, which arise from the effective time-reversal symmetry breaking induced by incident light. This results in optically switchable topological transition between FCTSC phases carrying different Chern numbers.
Our work provides an experimentally accessible approach for manipulating and designing topological superconducting states. Considering the magnet-superconductor hybrid systems with emergent topological superconductivity \cite{MHS1,MHS2,MHS3,MHS4,MHS5,MHS6,MHS7,MHS8} and especially the recent experimental realization of antiferromagnetic TNPSCs \cite{TNPSC}, our work will attracted wide attention for investigating Floquet engineering quantum states intertwining with magnetism, superconductivity, and topology in experiments.

\emph{{\color{magenta}Acknowledgments.}}---This work was supported by the National Natural Science Foundation of China (NSFC, Grants No. 12304191, No. 12222402, No. 92365101, No. 12074108, and 12347101), the Chongqing Natural Science Foundation (Grants No. CSTB2023NSCQ-JQX0024 and No. CSTB2022NSCQ-MSX0568), and  the Fundamental Research Funds for the Central Universities (Grant No. 2023CDJXY-048).


\begin{thebibliography}{76}%
\makeatletter
\providecommand \@ifxundefined [1]{%
 \@ifx{#1\undefined}
}%
\providecommand \@ifnum [1]{%
 \ifnum #1\expandafter \@firstoftwo
 \else \expandafter \@secondoftwo
 \fi
}%
\providecommand \@ifx [1]{%
 \ifx #1\expandafter \@firstoftwo
 \else \expandafter \@secondoftwo
 \fi
}%
\providecommand \natexlab [1]{#1}%
\providecommand \enquote  [1]{``#1''}%
\providecommand \bibnamefont  [1]{#1}%
\providecommand \bibfnamefont [1]{#1}%
\providecommand \citenamefont [1]{#1}%
\providecommand \href@noop [0]{\@secondoftwo}%
\providecommand \href [0]{\begingroup \@sanitize@url \@href}%
\providecommand \@href[1]{\@@startlink{#1}\@@href}%
\providecommand \@@href[1]{\endgroup#1\@@endlink}%
\providecommand \@sanitize@url [0]{\catcode `\\12\catcode `\$12\catcode
  `\&12\catcode `\#12\catcode `\^12\catcode `\_12\catcode `\%12\relax}%
\providecommand \@@startlink[1]{}%
\providecommand \@@endlink[0]{}%
\providecommand \url  [0]{\begingroup\@sanitize@url \@url }%
\providecommand \@url [1]{\endgroup\@href {#1}{\urlprefix }}%
\providecommand \urlprefix  [0]{URL }%
\providecommand \Eprint [0]{\href }%
\providecommand \doibase [0]{https://doi.org/}%
\providecommand \selectlanguage [0]{\@gobble}%
\providecommand \bibinfo  [0]{\@secondoftwo}%
\providecommand \bibfield  [0]{\@secondoftwo}%
\providecommand \translation [1]{[#1]}%
\providecommand \BibitemOpen [0]{}%
\providecommand \bibitemStop [0]{}%
\providecommand \bibitemNoStop [0]{.\EOS\space}%
\providecommand \EOS [0]{\spacefactor3000\relax}%
\providecommand \BibitemShut  [1]{\csname bibitem#1\endcsname}%
\let\auto@bib@innerbib\@empty
\bibitem [{\citenamefont {Hasan}\ and\ \citenamefont {Kane}(2010)}]{TI1}%
  \BibitemOpen
  \bibfield  {author} {\bibinfo {author} {\bibfnamefont {M.~Z.}\ \bibnamefont
  {Hasan}}\ and\ \bibinfo {author} {\bibfnamefont {C.~L.}\ \bibnamefont
  {Kane}},\ }\href {https://doi.org/10.1103/RevModPhys.82.3045} {\bibfield
  {journal} {\bibinfo  {journal} {Rev. Mod. Phys.}\ }\textbf {\bibinfo {volume}
  {82}},\ \bibinfo {pages} {3045} (\bibinfo {year} {2010})}\BibitemShut
  {NoStop}%
\bibitem [{\citenamefont {Qi}\ and\ \citenamefont {Zhang}(2011)}]{TI2}%
  \BibitemOpen
  \bibfield  {author} {\bibinfo {author} {\bibfnamefont {X.-L.}\ \bibnamefont
  {Qi}}\ and\ \bibinfo {author} {\bibfnamefont {S.-C.}\ \bibnamefont {Zhang}},\
  }\href {https://doi.org/10.1103/RevModPhys.83.1057} {\bibfield  {journal}
  {\bibinfo  {journal} {Rev. Mod. Phys.}\ }\textbf {\bibinfo {volume} {83}},\
  \bibinfo {pages} {1057} (\bibinfo {year} {2011})}\BibitemShut {NoStop}%
\bibitem [{\citenamefont {Kitaev}(2001)}]{MF1}%
  \BibitemOpen
  \bibfield  {author} {\bibinfo {author} {\bibfnamefont {A.~Y.}\ \bibnamefont
  {Kitaev}},\ }\href {https://doi.org/10.1070/1063-7869/44/10S/S29} {\bibfield
  {journal} {\bibinfo  {journal} {Phys. Usp.}\ }\textbf {\bibinfo {volume}
  {44}},\ \bibinfo {pages} {131} (\bibinfo {year} {2001})}\BibitemShut
  {NoStop}%
\bibitem [{\citenamefont {Fu}\ and\ \citenamefont {Kane}(2008)}]{MF2}%
  \BibitemOpen
  \bibfield  {author} {\bibinfo {author} {\bibfnamefont {L.}~\bibnamefont
  {Fu}}\ and\ \bibinfo {author} {\bibfnamefont {C.~L.}\ \bibnamefont {Kane}},\
  }\href {https://doi.org/10.1103/PhysRevLett.100.096407} {\bibfield  {journal}
  {\bibinfo  {journal} {Phys. Rev. Lett.}\ }\textbf {\bibinfo {volume} {100}},\
  \bibinfo {pages} {096407} (\bibinfo {year} {2008})}\BibitemShut {NoStop}%
\bibitem [{\citenamefont {Wilczek}(2009)}]{MF3}%
  \BibitemOpen
  \bibfield  {author} {\bibinfo {author} {\bibfnamefont {F.}~\bibnamefont
  {Wilczek}},\ }\href {https://www.nature.com/articles/nphys1380} {\bibfield
  {journal} {\bibinfo  {journal} {Nature Physics}\ }\textbf {\bibinfo {volume}
  {5}},\ \bibinfo {pages} {614} (\bibinfo {year} {2009})}\BibitemShut {NoStop}%
\bibitem [{\citenamefont {Lutchyn}\ \emph {et~al.}(2010)\citenamefont
  {Lutchyn}, \citenamefont {Sau},\ and\ \citenamefont {Das~Sarma}}]{MF4}%
  \BibitemOpen
  \bibfield  {author} {\bibinfo {author} {\bibfnamefont {R.~M.}\ \bibnamefont
  {Lutchyn}}, \bibinfo {author} {\bibfnamefont {J.~D.}\ \bibnamefont {Sau}},\
  and\ \bibinfo {author} {\bibfnamefont {S.}~\bibnamefont {Das~Sarma}},\ }\href
  {https://doi.org/10.1103/PhysRevLett.105.077001} {\bibfield  {journal}
  {\bibinfo  {journal} {Phys. Rev. Lett.}\ }\textbf {\bibinfo {volume} {105}},\
  \bibinfo {pages} {077001} (\bibinfo {year} {2010})}\BibitemShut {NoStop}%
\bibitem [{\citenamefont {Oreg}\ \emph {et~al.}(2010)\citenamefont {Oreg},
  \citenamefont {Refael},\ and\ \citenamefont {von Oppen}}]{MF5}%
  \BibitemOpen
  \bibfield  {author} {\bibinfo {author} {\bibfnamefont {Y.}~\bibnamefont
  {Oreg}}, \bibinfo {author} {\bibfnamefont {G.}~\bibnamefont {Refael}},\ and\
  \bibinfo {author} {\bibfnamefont {F.}~\bibnamefont {von Oppen}},\ }\href
  {https://doi.org/10.1103/PhysRevLett.105.177002} {\bibfield  {journal}
  {\bibinfo  {journal} {Phys. Rev. Lett.}\ }\textbf {\bibinfo {volume} {105}},\
  \bibinfo {pages} {177002} (\bibinfo {year} {2010})}\BibitemShut {NoStop}%
\bibitem [{\citenamefont {Beenakker}(2013)}]{MF6}%
  \BibitemOpen
  \bibfield  {author} {\bibinfo {author} {\bibfnamefont {C.}~\bibnamefont
  {Beenakker}},\ }\href
  {https://www.annualreviews.org/doi/abs/10.1146/annurev-conmatphys-030212-184337}
  {\bibfield  {journal} {\bibinfo  {journal} {Annu. Rev. Condens. Matter
  Phys.}\ }\textbf {\bibinfo {volume} {4}},\ \bibinfo {pages} {113} (\bibinfo
  {year} {2013})}\BibitemShut {NoStop}%
\bibitem [{\citenamefont {Ivanov}(2001)}]{Ivanov2001}%
  \BibitemOpen
  \bibfield  {author} {\bibinfo {author} {\bibfnamefont {D.~A.}\ \bibnamefont
  {Ivanov}},\ }\href {https://doi.org/10.1103/PhysRevLett.86.268} {\bibfield
  {journal} {\bibinfo  {journal} {Phys. Rev. Lett.}\ }\textbf {\bibinfo
  {volume} {86}},\ \bibinfo {pages} {268} (\bibinfo {year} {2001})}\BibitemShut
  {NoStop}%
\bibitem [{\citenamefont {Kitaev}(2003)}]{tqc1}%
  \BibitemOpen
  \bibfield  {author} {\bibinfo {author} {\bibfnamefont {A.}~\bibnamefont
  {Kitaev}},\ }\href
  {https://doi.org/https://doi.org/10.1016/S0003-4916(02)00018-0} {\bibfield
  {journal} {\bibinfo  {journal} {Ann. Phys.}\ }\textbf {\bibinfo {volume}
  {303}},\ \bibinfo {pages} {2} (\bibinfo {year} {2003})}\BibitemShut {NoStop}%
\bibitem [{\citenamefont {Sau}\ \emph {et~al.}(2010)\citenamefont {Sau},
  \citenamefont {Lutchyn}, \citenamefont {Tewari},\ and\ \citenamefont
  {Das~Sarma}}]{tqc2}%
  \BibitemOpen
  \bibfield  {author} {\bibinfo {author} {\bibfnamefont {J.~D.}\ \bibnamefont
  {Sau}}, \bibinfo {author} {\bibfnamefont {R.~M.}\ \bibnamefont {Lutchyn}},
  \bibinfo {author} {\bibfnamefont {S.}~\bibnamefont {Tewari}},\ and\ \bibinfo
  {author} {\bibfnamefont {S.}~\bibnamefont {Das~Sarma}},\ }\href
  {https://doi.org/10.1103/PhysRevLett.104.040502} {\bibfield  {journal}
  {\bibinfo  {journal} {Phys. Rev. Lett.}\ }\textbf {\bibinfo {volume} {104}},\
  \bibinfo {pages} {040502} (\bibinfo {year} {2010})}\BibitemShut {NoStop}%
\bibitem [{\citenamefont {Sarma}\ \emph {et~al.}(2015)\citenamefont {Sarma},
  \citenamefont {Freedman},\ and\ \citenamefont {Nayak}}]{tqc3}%
  \BibitemOpen
  \bibfield  {author} {\bibinfo {author} {\bibfnamefont {S.~D.}\ \bibnamefont
  {Sarma}}, \bibinfo {author} {\bibfnamefont {M.}~\bibnamefont {Freedman}},\
  and\ \bibinfo {author} {\bibfnamefont {C.}~\bibnamefont {Nayak}},\ }\href
  {https://www.nature.com/articles/npjqi20151} {\bibfield  {journal} {\bibinfo
  {journal} {npj Quantum Inf.}\ }\textbf {\bibinfo {volume} {1}},\ \bibinfo
  {pages} {1} (\bibinfo {year} {2015})}\BibitemShut {NoStop}%
\bibitem [{\citenamefont {Lutchyn}\ \emph {et~al.}(2018)\citenamefont
  {Lutchyn}, \citenamefont {Bakkers}, \citenamefont {Kouwenhoven},
  \citenamefont {Krogstrup}, \citenamefont {Marcus},\ and\ \citenamefont
  {Oreg}}]{tqc4}%
  \BibitemOpen
  \bibfield  {author} {\bibinfo {author} {\bibfnamefont {R.~M.}\ \bibnamefont
  {Lutchyn}}, \bibinfo {author} {\bibfnamefont {E.~P.}\ \bibnamefont
  {Bakkers}}, \bibinfo {author} {\bibfnamefont {L.~P.}\ \bibnamefont
  {Kouwenhoven}}, \bibinfo {author} {\bibfnamefont {P.}~\bibnamefont
  {Krogstrup}}, \bibinfo {author} {\bibfnamefont {C.~M.}\ \bibnamefont
  {Marcus}},\ and\ \bibinfo {author} {\bibfnamefont {Y.}~\bibnamefont {Oreg}},\
  }\href {https://www.nature.com/articles/s41578-018-0003-1} {\bibfield
  {journal} {\bibinfo  {journal} {Nat. Rev. Mater.}\ }\textbf {\bibinfo
  {volume} {3}},\ \bibinfo {pages} {52} (\bibinfo {year} {2018})}\BibitemShut
  {NoStop}%
\bibitem [{\citenamefont {Prada}\ \emph {et~al.}(2020)\citenamefont {Prada},
  \citenamefont {San-Jose}, \citenamefont {de~Moor}, \citenamefont {Geresdi},
  \citenamefont {Lee}, \citenamefont {Klinovaja}, \citenamefont {Loss},
  \citenamefont {Nyg{\aa}rd}, \citenamefont {Aguado},\ and\ \citenamefont
  {Kouwenhoven}}]{tqc5}%
  \BibitemOpen
  \bibfield  {author} {\bibinfo {author} {\bibfnamefont {E.}~\bibnamefont
  {Prada}}, \bibinfo {author} {\bibfnamefont {P.}~\bibnamefont {San-Jose}},
  \bibinfo {author} {\bibfnamefont {M.~W.}\ \bibnamefont {de~Moor}}, \bibinfo
  {author} {\bibfnamefont {A.}~\bibnamefont {Geresdi}}, \bibinfo {author}
  {\bibfnamefont {E.~J.}\ \bibnamefont {Lee}}, \bibinfo {author} {\bibfnamefont
  {J.}~\bibnamefont {Klinovaja}}, \bibinfo {author} {\bibfnamefont
  {D.}~\bibnamefont {Loss}}, \bibinfo {author} {\bibfnamefont {J.}~\bibnamefont
  {Nyg{\aa}rd}}, \bibinfo {author} {\bibfnamefont {R.}~\bibnamefont {Aguado}},\
  and\ \bibinfo {author} {\bibfnamefont {L.~P.}\ \bibnamefont {Kouwenhoven}},\
  }\href {https://www.nature.com/articles/s42254-020-0228-y} {\bibfield
  {journal} {\bibinfo  {journal} {Nat. Rev. Mater.}\ }\textbf {\bibinfo
  {volume} {2}},\ \bibinfo {pages} {575} (\bibinfo {year} {2020})}\BibitemShut
  {NoStop}%
\bibitem [{\citenamefont {Sato}(2006)}]{NSC1}%
  \BibitemOpen
  \bibfield  {author} {\bibinfo {author} {\bibfnamefont {M.}~\bibnamefont
  {Sato}},\ }\href {https://doi.org/10.1103/PhysRevB.73.214502} {\bibfield
  {journal} {\bibinfo  {journal} {Phys. Rev. B}\ }\textbf {\bibinfo {volume}
  {73}},\ \bibinfo {pages} {214502} (\bibinfo {year} {2006})}\BibitemShut
  {NoStop}%
\bibitem [{\citenamefont {B\'eri}(2010)}]{NSC2}%
  \BibitemOpen
  \bibfield  {author} {\bibinfo {author} {\bibfnamefont {B.}~\bibnamefont
  {B\'eri}},\ }\href {https://doi.org/10.1103/PhysRevB.81.134515} {\bibfield
  {journal} {\bibinfo  {journal} {Phys. Rev. B}\ }\textbf {\bibinfo {volume}
  {81}},\ \bibinfo {pages} {134515} (\bibinfo {year} {2010})}\BibitemShut
  {NoStop}%
\bibitem [{\citenamefont {Sato}\ and\ \citenamefont {Fujimoto}(2010)}]{NSC3}%
  \BibitemOpen
  \bibfield  {author} {\bibinfo {author} {\bibfnamefont {M.}~\bibnamefont
  {Sato}}\ and\ \bibinfo {author} {\bibfnamefont {S.}~\bibnamefont
  {Fujimoto}},\ }\href {https://doi.org/10.1103/PhysRevLett.105.217001}
  {\bibfield  {journal} {\bibinfo  {journal} {Phys. Rev. Lett.}\ }\textbf
  {\bibinfo {volume} {105}},\ \bibinfo {pages} {217001} (\bibinfo {year}
  {2010})}\BibitemShut {NoStop}%
\bibitem [{\citenamefont {Tanaka}\ \emph {et~al.}(2010)\citenamefont {Tanaka},
  \citenamefont {Mizuno}, \citenamefont {Yokoyama}, \citenamefont {Yada},\ and\
  \citenamefont {Sato}}]{NSC4}%
  \BibitemOpen
  \bibfield  {author} {\bibinfo {author} {\bibfnamefont {Y.}~\bibnamefont
  {Tanaka}}, \bibinfo {author} {\bibfnamefont {Y.}~\bibnamefont {Mizuno}},
  \bibinfo {author} {\bibfnamefont {T.}~\bibnamefont {Yokoyama}}, \bibinfo
  {author} {\bibfnamefont {K.}~\bibnamefont {Yada}},\ and\ \bibinfo {author}
  {\bibfnamefont {M.}~\bibnamefont {Sato}},\ }\href
  {https://doi.org/10.1103/PhysRevLett.105.097002} {\bibfield  {journal}
  {\bibinfo  {journal} {Phys. Rev. Lett.}\ }\textbf {\bibinfo {volume} {105}},\
  \bibinfo {pages} {097002} (\bibinfo {year} {2010})}\BibitemShut {NoStop}%
\bibitem [{\citenamefont {Sato}\ \emph {et~al.}(2011)\citenamefont {Sato},
  \citenamefont {Tanaka}, \citenamefont {Yada},\ and\ \citenamefont
  {Yokoyama}}]{NSC5}%
  \BibitemOpen
  \bibfield  {author} {\bibinfo {author} {\bibfnamefont {M.}~\bibnamefont
  {Sato}}, \bibinfo {author} {\bibfnamefont {Y.}~\bibnamefont {Tanaka}},
  \bibinfo {author} {\bibfnamefont {K.}~\bibnamefont {Yada}},\ and\ \bibinfo
  {author} {\bibfnamefont {T.}~\bibnamefont {Yokoyama}},\ }\href
  {https://doi.org/10.1103/PhysRevB.83.224511} {\bibfield  {journal} {\bibinfo
  {journal} {Phys. Rev. B}\ }\textbf {\bibinfo {volume} {83}},\ \bibinfo
  {pages} {224511} (\bibinfo {year} {2011})}\BibitemShut {NoStop}%
\bibitem [{\citenamefont {Deng}\ \emph {et~al.}(2014)\citenamefont {Deng},
  \citenamefont {Ortiz}, \citenamefont {Poudel},\ and\ \citenamefont
  {Viola}}]{NSC6}%
  \BibitemOpen
  \bibfield  {author} {\bibinfo {author} {\bibfnamefont {S.}~\bibnamefont
  {Deng}}, \bibinfo {author} {\bibfnamefont {G.}~\bibnamefont {Ortiz}},
  \bibinfo {author} {\bibfnamefont {A.}~\bibnamefont {Poudel}},\ and\ \bibinfo
  {author} {\bibfnamefont {L.}~\bibnamefont {Viola}},\ }\href
  {https://doi.org/10.1103/PhysRevB.89.140507} {\bibfield  {journal} {\bibinfo
  {journal} {Phys. Rev. B}\ }\textbf {\bibinfo {volume} {89}},\ \bibinfo
  {pages} {140507} (\bibinfo {year} {2014})}\BibitemShut {NoStop}%
\bibitem [{\citenamefont {Wong}\ \emph {et~al.}(2013)\citenamefont {Wong},
  \citenamefont {Liu}, \citenamefont {Law},\ and\ \citenamefont {Lee}}]{NSC7}%
  \BibitemOpen
  \bibfield  {author} {\bibinfo {author} {\bibfnamefont {C.~L.~M.}\
  \bibnamefont {Wong}}, \bibinfo {author} {\bibfnamefont {J.}~\bibnamefont
  {Liu}}, \bibinfo {author} {\bibfnamefont {K.~T.}\ \bibnamefont {Law}},\ and\
  \bibinfo {author} {\bibfnamefont {P.~A.}\ \bibnamefont {Lee}},\ }\href
  {https://doi.org/10.1103/PhysRevB.88.060504} {\bibfield  {journal} {\bibinfo
  {journal} {Phys. Rev. B}\ }\textbf {\bibinfo {volume} {88}},\ \bibinfo
  {pages} {060504} (\bibinfo {year} {2013})}\BibitemShut {NoStop}%
\bibitem [{\citenamefont {Kao}\ \emph {et~al.}(2015)\citenamefont {Kao},
  \citenamefont {Huang}, \citenamefont {Mou},\ and\ \citenamefont
  {Tsuei}}]{NSC8}%
  \BibitemOpen
  \bibfield  {author} {\bibinfo {author} {\bibfnamefont {J.-T.}\ \bibnamefont
  {Kao}}, \bibinfo {author} {\bibfnamefont {S.-M.}\ \bibnamefont {Huang}},
  \bibinfo {author} {\bibfnamefont {C.-Y.}\ \bibnamefont {Mou}},\ and\ \bibinfo
  {author} {\bibfnamefont {C.~C.}\ \bibnamefont {Tsuei}},\ }\href
  {https://doi.org/10.1103/PhysRevB.91.134501} {\bibfield  {journal} {\bibinfo
  {journal} {Phys. Rev. B}\ }\textbf {\bibinfo {volume} {91}},\ \bibinfo
  {pages} {134501} (\bibinfo {year} {2015})}\BibitemShut {NoStop}%
\bibitem [{\citenamefont {Baum}\ \emph {et~al.}(2015)\citenamefont {Baum},
  \citenamefont {Posske}, \citenamefont {Fulga}, \citenamefont {Trauzettel},\
  and\ \citenamefont {Stern}}]{NSC9}%
  \BibitemOpen
  \bibfield  {author} {\bibinfo {author} {\bibfnamefont {Y.}~\bibnamefont
  {Baum}}, \bibinfo {author} {\bibfnamefont {T.}~\bibnamefont {Posske}},
  \bibinfo {author} {\bibfnamefont {I.~C.}\ \bibnamefont {Fulga}}, \bibinfo
  {author} {\bibfnamefont {B.}~\bibnamefont {Trauzettel}},\ and\ \bibinfo
  {author} {\bibfnamefont {A.}~\bibnamefont {Stern}},\ }\href
  {https://doi.org/10.1103/PhysRevB.92.045128} {\bibfield  {journal} {\bibinfo
  {journal} {Phys. Rev. B}\ }\textbf {\bibinfo {volume} {92}},\ \bibinfo
  {pages} {045128} (\bibinfo {year} {2015})}\BibitemShut {NoStop}%
\bibitem [{\citenamefont {Schnyder}\ and\ \citenamefont
  {Brydon}(2015)}]{NSC10}%
  \BibitemOpen
  \bibfield  {author} {\bibinfo {author} {\bibfnamefont {A.~P.}\ \bibnamefont
  {Schnyder}}\ and\ \bibinfo {author} {\bibfnamefont {P.~M.~R.}\ \bibnamefont
  {Brydon}},\ }\href {https://doi.org/10.1088/0953-8984/27/24/243201}
  {\bibfield  {journal} {\bibinfo  {journal} {J. Phys. Condens. Matter}\
  }\textbf {\bibinfo {volume} {27}},\ \bibinfo {pages} {243201} (\bibinfo
  {year} {2015})}\BibitemShut {NoStop}%
\bibitem [{\citenamefont {Brzezicki}\ and\ \citenamefont
  {Cuoco}(2018)}]{NSC11}%
  \BibitemOpen
  \bibfield  {author} {\bibinfo {author} {\bibfnamefont {W.}~\bibnamefont
  {Brzezicki}}\ and\ \bibinfo {author} {\bibfnamefont {M.}~\bibnamefont
  {Cuoco}},\ }\href {https://doi.org/10.1103/PhysRevB.97.064513} {\bibfield
  {journal} {\bibinfo  {journal} {Phys. Rev. B}\ }\textbf {\bibinfo {volume}
  {97}},\ \bibinfo {pages} {064513} (\bibinfo {year} {2018})}\BibitemShut
  {NoStop}%
\bibitem [{\citenamefont {Bouhon}\ \emph {et~al.}(2018)\citenamefont {Bouhon},
  \citenamefont {Schmidt},\ and\ \citenamefont {Black-Schaffer}}]{NSC12}%
  \BibitemOpen
  \bibfield  {author} {\bibinfo {author} {\bibfnamefont {A.}~\bibnamefont
  {Bouhon}}, \bibinfo {author} {\bibfnamefont {J.}~\bibnamefont {Schmidt}},\
  and\ \bibinfo {author} {\bibfnamefont {A.~M.}\ \bibnamefont
  {Black-Schaffer}},\ }\href {https://doi.org/10.1103/PhysRevB.97.104508}
  {\bibfield  {journal} {\bibinfo  {journal} {Phys. Rev. B}\ }\textbf {\bibinfo
  {volume} {97}},\ \bibinfo {pages} {104508} (\bibinfo {year}
  {2018})}\BibitemShut {NoStop}%
\bibitem [{\citenamefont {Zhang}\ \emph {et~al.}(2019)\citenamefont {Zhang},
  \citenamefont {Cole}, \citenamefont {Wu},\ and\ \citenamefont
  {Das~Sarma}}]{NSC13}%
  \BibitemOpen
  \bibfield  {author} {\bibinfo {author} {\bibfnamefont {R.-X.}\ \bibnamefont
  {Zhang}}, \bibinfo {author} {\bibfnamefont {W.~S.}\ \bibnamefont {Cole}},
  \bibinfo {author} {\bibfnamefont {X.}~\bibnamefont {Wu}},\ and\ \bibinfo
  {author} {\bibfnamefont {S.}~\bibnamefont {Das~Sarma}},\ }\href
  {https://doi.org/10.1103/PhysRevLett.123.167001} {\bibfield  {journal}
  {\bibinfo  {journal} {Phys. Rev. Lett.}\ }\textbf {\bibinfo {volume} {123}},\
  \bibinfo {pages} {167001} (\bibinfo {year} {2019})}\BibitemShut {NoStop}%
\bibitem [{\citenamefont {Wang}\ \emph {et~al.}(2018)\citenamefont {Wang},
  \citenamefont {Rosdahl},\ and\ \citenamefont {Sticlet}}]{NSC14}%
  \BibitemOpen
  \bibfield  {author} {\bibinfo {author} {\bibfnamefont {L.}~\bibnamefont
  {Wang}}, \bibinfo {author} {\bibfnamefont {T.~O.}\ \bibnamefont {Rosdahl}},\
  and\ \bibinfo {author} {\bibfnamefont {D.}~\bibnamefont {Sticlet}},\ }\href
  {https://doi.org/10.1103/PhysRevB.98.205411} {\bibfield  {journal} {\bibinfo
  {journal} {Phys. Rev. B}\ }\textbf {\bibinfo {volume} {98}},\ \bibinfo
  {pages} {205411} (\bibinfo {year} {2018})}\BibitemShut {NoStop}%
\bibitem [{\citenamefont {Nayak}\ \emph {et~al.}(2021)\citenamefont {Nayak},
  \citenamefont {Steinbok}, \citenamefont {Roet}, \citenamefont {Koo},
  \citenamefont {Margalit}, \citenamefont {Feldman}, \citenamefont {Almoalem},
  \citenamefont {Kanigel}, \citenamefont {Fiete}, \citenamefont {Yan} \emph
  {et~al.}}]{NSC15}%
  \BibitemOpen
  \bibfield  {author} {\bibinfo {author} {\bibfnamefont {A.~K.}\ \bibnamefont
  {Nayak}}, \bibinfo {author} {\bibfnamefont {A.}~\bibnamefont {Steinbok}},
  \bibinfo {author} {\bibfnamefont {Y.}~\bibnamefont {Roet}}, \bibinfo {author}
  {\bibfnamefont {J.}~\bibnamefont {Koo}}, \bibinfo {author} {\bibfnamefont
  {G.}~\bibnamefont {Margalit}}, \bibinfo {author} {\bibfnamefont
  {I.}~\bibnamefont {Feldman}}, \bibinfo {author} {\bibfnamefont
  {A.}~\bibnamefont {Almoalem}}, \bibinfo {author} {\bibfnamefont
  {A.}~\bibnamefont {Kanigel}}, \bibinfo {author} {\bibfnamefont {G.~A.}\
  \bibnamefont {Fiete}}, \bibinfo {author} {\bibfnamefont {B.}~\bibnamefont
  {Yan}}, \emph {et~al.},\ }\href
  {https://www.nature.com/articles/s41567-021-01376-z} {\bibfield  {journal}
  {\bibinfo  {journal} {Nat. Phys.}\ }\textbf {\bibinfo {volume} {17}},\
  \bibinfo {pages} {1413} (\bibinfo {year} {2021})}\BibitemShut {NoStop}%
\bibitem [{\citenamefont {Lu}\ \emph {et~al.}(2023)\citenamefont {Lu},
  \citenamefont {Wang}, \citenamefont {Zhang}, \citenamefont {Chen},
  \citenamefont {Cui},\ and\ \citenamefont {Luo}}]{NSC16}%
  \BibitemOpen
  \bibfield  {author} {\bibinfo {author} {\bibfnamefont {M.-L.}\ \bibnamefont
  {Lu}}, \bibinfo {author} {\bibfnamefont {Y.}~\bibnamefont {Wang}}, \bibinfo
  {author} {\bibfnamefont {H.-Z.}\ \bibnamefont {Zhang}}, \bibinfo {author}
  {\bibfnamefont {H.-L.}\ \bibnamefont {Chen}}, \bibinfo {author}
  {\bibfnamefont {T.-Y.}\ \bibnamefont {Cui}},\ and\ \bibinfo {author}
  {\bibfnamefont {X.}~\bibnamefont {Luo}},\ }\href
  {https://iopscience.iop.org/article/10.1088/1674-1056/ac7208/meta} {\bibfield
   {journal} {\bibinfo  {journal} {Chin. Phys. B}\ }\textbf {\bibinfo {volume}
  {32}},\ \bibinfo {pages} {027301} (\bibinfo {year} {2023})}\BibitemShut
  {NoStop}%
\bibitem [{\citenamefont {Chiu}\ \emph {et~al.}(2016)\citenamefont {Chiu},
  \citenamefont {Teo}, \citenamefont {Schnyder},\ and\ \citenamefont
  {Ryu}}]{RevModPhys2016}%
  \BibitemOpen
  \bibfield  {author} {\bibinfo {author} {\bibfnamefont {C.-K.}\ \bibnamefont
  {Chiu}}, \bibinfo {author} {\bibfnamefont {J.~C.~Y.}\ \bibnamefont {Teo}},
  \bibinfo {author} {\bibfnamefont {A.~P.}\ \bibnamefont {Schnyder}},\ and\
  \bibinfo {author} {\bibfnamefont {S.}~\bibnamefont {Ryu}},\ }\href
  {https://doi.org/10.1103/RevModPhys.88.035005} {\bibfield  {journal}
  {\bibinfo  {journal} {Rev. Mod. Phys.}\ }\textbf {\bibinfo {volume} {88}},\
  \bibinfo {pages} {035005} (\bibinfo {year} {2016})}\BibitemShut {NoStop}%
\bibitem [{\citenamefont {Shirley}(1965)}]{Shirley}%
  \BibitemOpen
  \bibfield  {author} {\bibinfo {author} {\bibfnamefont {J.~H.}\ \bibnamefont
  {Shirley}},\ }\href {https://doi.org/10.1103/PhysRev.138.B979} {\bibfield
  {journal} {\bibinfo  {journal} {Phys. Rev.}\ }\textbf {\bibinfo {volume}
  {138}},\ \bibinfo {pages} {B979} (\bibinfo {year} {1965})}\BibitemShut
  {NoStop}%
\bibitem [{\citenamefont {Goldman}\ and\ \citenamefont
  {Dalibard}(2014)}]{Goldman}%
  \BibitemOpen
  \bibfield  {author} {\bibinfo {author} {\bibfnamefont {N.}~\bibnamefont
  {Goldman}}\ and\ \bibinfo {author} {\bibfnamefont {J.}~\bibnamefont
  {Dalibard}},\ }\href {https://doi.org/10.1103/PhysRevX.4.031027} {\bibfield
  {journal} {\bibinfo  {journal} {Phys. Rev. X}\ }\textbf {\bibinfo {volume}
  {4}},\ \bibinfo {pages} {031027} (\bibinfo {year} {2014})}\BibitemShut
  {NoStop}%
\bibitem [{\citenamefont {Eckardt}\ and\ \citenamefont
  {Anisimovas}(2015)}]{Eckardt2015}%
  \BibitemOpen
  \bibfield  {author} {\bibinfo {author} {\bibfnamefont {A.}~\bibnamefont
  {Eckardt}}\ and\ \bibinfo {author} {\bibfnamefont {E.}~\bibnamefont
  {Anisimovas}},\ }\href {https://doi.org/10.1088/1367-2630/17/9/093039}
  {\bibfield  {journal} {\bibinfo  {journal} {New J. Phys.}\ }\textbf {\bibinfo
  {volume} {17}},\ \bibinfo {pages} {093039} (\bibinfo {year}
  {2015})}\BibitemShut {NoStop}%
\bibitem [{\citenamefont {Bukov}\ \emph {et~al.}(2015)\citenamefont {Bukov},
  \citenamefont {D'Alessio},\ and\ \citenamefont {Polkovnikov}}]{Bukov2015}%
  \BibitemOpen
  \bibfield  {author} {\bibinfo {author} {\bibfnamefont {M.}~\bibnamefont
  {Bukov}}, \bibinfo {author} {\bibfnamefont {L.}~\bibnamefont {D'Alessio}},\
  and\ \bibinfo {author} {\bibfnamefont {A.}~\bibnamefont {Polkovnikov}},\
  }\href {https://doi.org/10.1080/00018732.2015.1055918} {\bibfield  {journal}
  {\bibinfo  {journal} {Adv. Phys.}\ }\textbf {\bibinfo {volume} {64}},\
  \bibinfo {pages} {139} (\bibinfo {year} {2015})}\BibitemShut {NoStop}%
\bibitem [{\citenamefont {Oka}\ and\ \citenamefont {Kitamura}(2019)}]{Oka2019}%
  \BibitemOpen
  \bibfield  {author} {\bibinfo {author} {\bibfnamefont {T.}~\bibnamefont
  {Oka}}\ and\ \bibinfo {author} {\bibfnamefont {S.}~\bibnamefont {Kitamura}},\
  }\href {https://doi.org/10.1146/annurev-conmatphys-031218-013423} {\bibfield
  {journal} {\bibinfo  {journal} {Annu. Rev. Condens. Matter Phys.}\ }\textbf
  {\bibinfo {volume} {10}},\ \bibinfo {pages} {387} (\bibinfo {year}
  {2019})}\BibitemShut {NoStop}%
\bibitem [{\citenamefont {Wang}\ \emph {et~al.}(2013)\citenamefont {Wang},
  \citenamefont {Steinberg}, \citenamefont {Jarillo-Herrero},\ and\
  \citenamefont {Gedik}}]{exp1}%
  \BibitemOpen
  \bibfield  {author} {\bibinfo {author} {\bibfnamefont {Y.}~\bibnamefont
  {Wang}}, \bibinfo {author} {\bibfnamefont {H.}~\bibnamefont {Steinberg}},
  \bibinfo {author} {\bibfnamefont {P.}~\bibnamefont {Jarillo-Herrero}},\ and\
  \bibinfo {author} {\bibfnamefont {N.}~\bibnamefont {Gedik}},\ }\href
  {https://doi.org/10.1126/science.1239834} {\bibfield  {journal} {\bibinfo
  {journal} {Science}\ }\textbf {\bibinfo {volume} {342}},\ \bibinfo {pages}
  {453} (\bibinfo {year} {2013})}\BibitemShut {NoStop}%
\bibitem [{\citenamefont {Mahmood}\ \emph {et~al.}(2016)\citenamefont
  {Mahmood}, \citenamefont {Chan}, \citenamefont {Alpichshev}, \citenamefont
  {Gardner}, \citenamefont {Lee}, \citenamefont {Lee},\ and\ \citenamefont
  {Gedik}}]{exp2}%
  \BibitemOpen
  \bibfield  {author} {\bibinfo {author} {\bibfnamefont {F.}~\bibnamefont
  {Mahmood}}, \bibinfo {author} {\bibfnamefont {C.-K.}\ \bibnamefont {Chan}},
  \bibinfo {author} {\bibfnamefont {Z.}~\bibnamefont {Alpichshev}}, \bibinfo
  {author} {\bibfnamefont {D.}~\bibnamefont {Gardner}}, \bibinfo {author}
  {\bibfnamefont {Y.}~\bibnamefont {Lee}}, \bibinfo {author} {\bibfnamefont
  {P.~A.}\ \bibnamefont {Lee}},\ and\ \bibinfo {author} {\bibfnamefont
  {N.}~\bibnamefont {Gedik}},\ }\href
  {https://doi.org/https://doi.org/10.1038/nphys3609} {\bibfield  {journal}
  {\bibinfo  {journal} {Nat. Phys.}\ }\textbf {\bibinfo {volume} {12}},\
  \bibinfo {pages} {306} (\bibinfo {year} {2016})}\BibitemShut {NoStop}%
\bibitem [{\citenamefont {McIver}\ \emph {et~al.}(2020)\citenamefont {McIver},
  \citenamefont {Schulte}, \citenamefont {Stein}, \citenamefont {Matsuyama},
  \citenamefont {Jotzu}, \citenamefont {Meier},\ and\ \citenamefont
  {Cavalleri}}]{exp3}%
  \BibitemOpen
  \bibfield  {author} {\bibinfo {author} {\bibfnamefont {J.~W.}\ \bibnamefont
  {McIver}}, \bibinfo {author} {\bibfnamefont {B.}~\bibnamefont {Schulte}},
  \bibinfo {author} {\bibfnamefont {F.-U.}\ \bibnamefont {Stein}}, \bibinfo
  {author} {\bibfnamefont {T.}~\bibnamefont {Matsuyama}}, \bibinfo {author}
  {\bibfnamefont {G.}~\bibnamefont {Jotzu}}, \bibinfo {author} {\bibfnamefont
  {G.}~\bibnamefont {Meier}},\ and\ \bibinfo {author} {\bibfnamefont
  {A.}~\bibnamefont {Cavalleri}},\ }\href
  {https://doi.org/https://doi.org/10.1038/s41567-019-0698-y} {\bibfield
  {journal} {\bibinfo  {journal} {Nat. Phys.}\ }\textbf {\bibinfo {volume}
  {16}},\ \bibinfo {pages} {38} (\bibinfo {year} {2020})}\BibitemShut {NoStop}%
\bibitem [{\citenamefont {Harper}\ \emph {et~al.}(2020)\citenamefont {Harper},
  \citenamefont {Roy}, \citenamefont {Rudner},\ and\ \citenamefont
  {Sondhi}}]{Rev1}%
  \BibitemOpen
  \bibfield  {author} {\bibinfo {author} {\bibfnamefont {F.}~\bibnamefont
  {Harper}}, \bibinfo {author} {\bibfnamefont {R.}~\bibnamefont {Roy}},
  \bibinfo {author} {\bibfnamefont {M.~S.}\ \bibnamefont {Rudner}},\ and\
  \bibinfo {author} {\bibfnamefont {S.}~\bibnamefont {Sondhi}},\ }\href
  {https://doi.org/10.1146/annurev-conmatphys-031218-013721} {\bibfield
  {journal} {\bibinfo  {journal} {Annu. Rev. Condens. Matter Phys.}\ }\textbf
  {\bibinfo {volume} {11}},\ \bibinfo {pages} {345} (\bibinfo {year}
  {2020})}\BibitemShut {NoStop}%
\bibitem [{\citenamefont {Rudner}\ and\ \citenamefont {Lindner}(2020)}]{Rev2}%
  \BibitemOpen
  \bibfield  {author} {\bibinfo {author} {\bibfnamefont {M.~S.}\ \bibnamefont
  {Rudner}}\ and\ \bibinfo {author} {\bibfnamefont {N.~H.}\ \bibnamefont
  {Lindner}},\ }\href {https://doi.org/10.1038/s42254-020-0170-z} {\bibfield
  {journal} {\bibinfo  {journal} {Nat. Rev. Phys.}\ }\textbf {\bibinfo {volume}
  {2}},\ \bibinfo {pages} {229} (\bibinfo {year} {2020})}\BibitemShut {NoStop}%
\bibitem [{\citenamefont {Bao}\ \emph {et~al.}(2021)\citenamefont {Bao},
  \citenamefont {Tang}, \citenamefont {Sun},\ and\ \citenamefont
  {Zhou}}]{Rev3}%
  \BibitemOpen
  \bibfield  {author} {\bibinfo {author} {\bibfnamefont {C.}~\bibnamefont
  {Bao}}, \bibinfo {author} {\bibfnamefont {P.}~\bibnamefont {Tang}}, \bibinfo
  {author} {\bibfnamefont {D.}~\bibnamefont {Sun}},\ and\ \bibinfo {author}
  {\bibfnamefont {S.}~\bibnamefont {Zhou}},\ }\href
  {https://doi.org/10.1038/s42254-021-00388-1} {\bibfield  {journal} {\bibinfo
  {journal} {Nat. Rev. Phys.}\ ,\ \bibinfo {pages} {1}} (\bibinfo {year}
  {2021})}\BibitemShut {NoStop}%
\bibitem [{\citenamefont {Oka}\ and\ \citenamefont {Aoki}(2009)}]{FT1}%
  \BibitemOpen
  \bibfield  {author} {\bibinfo {author} {\bibfnamefont {T.}~\bibnamefont
  {Oka}}\ and\ \bibinfo {author} {\bibfnamefont {H.}~\bibnamefont {Aoki}},\
  }\href {https://doi.org/10.1103/PhysRevB.79.081406} {\bibfield  {journal}
  {\bibinfo  {journal} {Phys. Rev. B}\ }\textbf {\bibinfo {volume} {79}},\
  \bibinfo {pages} {081406} (\bibinfo {year} {2009})}\BibitemShut {NoStop}%
\bibitem [{\citenamefont {Lindner}\ \emph {et~al.}(2011)\citenamefont
  {Lindner}, \citenamefont {Refael},\ and\ \citenamefont {Galitski}}]{FT2}%
  \BibitemOpen
  \bibfield  {author} {\bibinfo {author} {\bibfnamefont {N.~H.}\ \bibnamefont
  {Lindner}}, \bibinfo {author} {\bibfnamefont {G.}~\bibnamefont {Refael}},\
  and\ \bibinfo {author} {\bibfnamefont {V.}~\bibnamefont {Galitski}},\ }\href
  {https://doi.org/https://doi.org/10.1038/nphys1926} {\bibfield  {journal}
  {\bibinfo  {journal} {Nat. Phys.}\ }\textbf {\bibinfo {volume} {7}},\
  \bibinfo {pages} {490} (\bibinfo {year} {2011})}\BibitemShut {NoStop}%
\bibitem [{\citenamefont {H{\"u}bener}\ \emph {et~al.}(2017)\citenamefont
  {H{\"u}bener}, \citenamefont {Sentef}, \citenamefont {De~Giovannini},
  \citenamefont {Kemper},\ and\ \citenamefont {Rubio}}]{FSM1}%
  \BibitemOpen
  \bibfield  {author} {\bibinfo {author} {\bibfnamefont {H.}~\bibnamefont
  {H{\"u}bener}}, \bibinfo {author} {\bibfnamefont {M.~A.}\ \bibnamefont
  {Sentef}}, \bibinfo {author} {\bibfnamefont {U.}~\bibnamefont
  {De~Giovannini}}, \bibinfo {author} {\bibfnamefont {A.~F.}\ \bibnamefont
  {Kemper}},\ and\ \bibinfo {author} {\bibfnamefont {A.}~\bibnamefont
  {Rubio}},\ }\href {https://doi.org/https://doi.org/10.1038/ncomms13940}
  {\bibfield  {journal} {\bibinfo  {journal} {Nat. Commun.}\ }\textbf {\bibinfo
  {volume} {8}},\ \bibinfo {pages} {13940} (\bibinfo {year}
  {2017})}\BibitemShut {NoStop}%
\bibitem [{\citenamefont {Yan}\ and\ \citenamefont {Wang}(2016)}]{FSM2}%
  \BibitemOpen
  \bibfield  {author} {\bibinfo {author} {\bibfnamefont {Z.}~\bibnamefont
  {Yan}}\ and\ \bibinfo {author} {\bibfnamefont {Z.}~\bibnamefont {Wang}},\
  }\href {https://doi.org/10.1103/PhysRevLett.117.087402} {\bibfield  {journal}
  {\bibinfo  {journal} {Phys. Rev. Lett.}\ }\textbf {\bibinfo {volume} {117}},\
  \bibinfo {pages} {087402} (\bibinfo {year} {2016})}\BibitemShut {NoStop}%
\bibitem [{\citenamefont {Gonz\'alez}\ and\ \citenamefont
  {Molina}(2016)}]{FSM3}%
  \BibitemOpen
  \bibfield  {author} {\bibinfo {author} {\bibfnamefont {J.}~\bibnamefont
  {Gonz\'alez}}\ and\ \bibinfo {author} {\bibfnamefont {R.~A.}\ \bibnamefont
  {Molina}},\ }\href {https://doi.org/10.1103/PhysRevLett.116.156803}
  {\bibfield  {journal} {\bibinfo  {journal} {Phys. Rev. Lett.}\ }\textbf
  {\bibinfo {volume} {116}},\ \bibinfo {pages} {156803} (\bibinfo {year}
  {2016})}\BibitemShut {NoStop}%
\bibitem [{\citenamefont {Trevisan}\ \emph {et~al.}(2022)\citenamefont
  {Trevisan}, \citenamefont {Arribi}, \citenamefont {Heinonen}, \citenamefont
  {Slager},\ and\ \citenamefont {Orth}}]{FSM4}%
  \BibitemOpen
  \bibfield  {author} {\bibinfo {author} {\bibfnamefont {T.~V.}\ \bibnamefont
  {Trevisan}}, \bibinfo {author} {\bibfnamefont {P.~V.}\ \bibnamefont
  {Arribi}}, \bibinfo {author} {\bibfnamefont {O.}~\bibnamefont {Heinonen}},
  \bibinfo {author} {\bibfnamefont {R.-J.}\ \bibnamefont {Slager}},\ and\
  \bibinfo {author} {\bibfnamefont {P.~P.}\ \bibnamefont {Orth}},\ }\href
  {https://doi.org/10.1103/PhysRevLett.128.066602} {\bibfield  {journal}
  {\bibinfo  {journal} {Phys. Rev. Lett.}\ }\textbf {\bibinfo {volume} {128}},\
  \bibinfo {pages} {066602} (\bibinfo {year} {2022})}\BibitemShut {NoStop}%
\bibitem [{\citenamefont {Lu}\ \emph {et~al.}(2022)\citenamefont {Lu},
  \citenamefont {Reid}, \citenamefont {Fritsch}, \citenamefont {Pi\~neiro},\
  and\ \citenamefont {Spielman}}]{FSM5}%
  \BibitemOpen
  \bibfield  {author} {\bibinfo {author} {\bibfnamefont {M.}~\bibnamefont
  {Lu}}, \bibinfo {author} {\bibfnamefont {G.~H.}\ \bibnamefont {Reid}},
  \bibinfo {author} {\bibfnamefont {A.~R.}\ \bibnamefont {Fritsch}}, \bibinfo
  {author} {\bibfnamefont {A.~M.}\ \bibnamefont {Pi\~neiro}},\ and\ \bibinfo
  {author} {\bibfnamefont {I.~B.}\ \bibnamefont {Spielman}},\ }\href
  {https://doi.org/10.1103/PhysRevLett.129.040402} {\bibfield  {journal}
  {\bibinfo  {journal} {Phys. Rev. Lett.}\ }\textbf {\bibinfo {volume} {129}},\
  \bibinfo {pages} {040402} (\bibinfo {year} {2022})}\BibitemShut {NoStop}%
\bibitem [{\citenamefont {Rudner}\ \emph {et~al.}(2013)\citenamefont {Rudner},
  \citenamefont {Lindner}, \citenamefont {Berg},\ and\ \citenamefont
  {Levin}}]{AES}%
  \BibitemOpen
  \bibfield  {author} {\bibinfo {author} {\bibfnamefont {M.~S.}\ \bibnamefont
  {Rudner}}, \bibinfo {author} {\bibfnamefont {N.~H.}\ \bibnamefont {Lindner}},
  \bibinfo {author} {\bibfnamefont {E.}~\bibnamefont {Berg}},\ and\ \bibinfo
  {author} {\bibfnamefont {M.}~\bibnamefont {Levin}},\ }\href
  {https://doi.org/10.1103/PhysRevX.3.031005} {\bibfield  {journal} {\bibinfo
  {journal} {Phys. Rev. X}\ }\textbf {\bibinfo {volume} {3}},\ \bibinfo {pages}
  {031005} (\bibinfo {year} {2013})}\BibitemShut {NoStop}%
\bibitem [{\citenamefont {Jiang}\ \emph {et~al.}(2011)\citenamefont {Jiang},
  \citenamefont {Kitagawa}, \citenamefont {Alicea}, \citenamefont {Akhmerov},
  \citenamefont {Pekker}, \citenamefont {Refael}, \citenamefont {Cirac},
  \citenamefont {Demler}, \citenamefont {Lukin},\ and\ \citenamefont
  {Zoller}}]{FMaj1}%
  \BibitemOpen
  \bibfield  {author} {\bibinfo {author} {\bibfnamefont {L.}~\bibnamefont
  {Jiang}}, \bibinfo {author} {\bibfnamefont {T.}~\bibnamefont {Kitagawa}},
  \bibinfo {author} {\bibfnamefont {J.}~\bibnamefont {Alicea}}, \bibinfo
  {author} {\bibfnamefont {A.~R.}\ \bibnamefont {Akhmerov}}, \bibinfo {author}
  {\bibfnamefont {D.}~\bibnamefont {Pekker}}, \bibinfo {author} {\bibfnamefont
  {G.}~\bibnamefont {Refael}}, \bibinfo {author} {\bibfnamefont {J.~I.}\
  \bibnamefont {Cirac}}, \bibinfo {author} {\bibfnamefont {E.}~\bibnamefont
  {Demler}}, \bibinfo {author} {\bibfnamefont {M.~D.}\ \bibnamefont {Lukin}},\
  and\ \bibinfo {author} {\bibfnamefont {P.}~\bibnamefont {Zoller}},\ }\href
  {https://doi.org/10.1103/PhysRevLett.106.220402} {\bibfield  {journal}
  {\bibinfo  {journal} {Phys. Rev. Lett.}\ }\textbf {\bibinfo {volume} {106}},\
  \bibinfo {pages} {220402} (\bibinfo {year} {2011})}\BibitemShut {NoStop}%
\bibitem [{\citenamefont {Benito}\ \emph {et~al.}(2014)\citenamefont {Benito},
  \citenamefont {G\'omez-Le\'on}, \citenamefont {Bastidas}, \citenamefont
  {Brandes},\ and\ \citenamefont {Platero}}]{FMaj2}%
  \BibitemOpen
  \bibfield  {author} {\bibinfo {author} {\bibfnamefont {M.}~\bibnamefont
  {Benito}}, \bibinfo {author} {\bibfnamefont {A.}~\bibnamefont
  {G\'omez-Le\'on}}, \bibinfo {author} {\bibfnamefont {V.~M.}\ \bibnamefont
  {Bastidas}}, \bibinfo {author} {\bibfnamefont {T.}~\bibnamefont {Brandes}},\
  and\ \bibinfo {author} {\bibfnamefont {G.}~\bibnamefont {Platero}},\ }\href
  {https://doi.org/10.1103/PhysRevB.90.205127} {\bibfield  {journal} {\bibinfo
  {journal} {Phys. Rev. B}\ }\textbf {\bibinfo {volume} {90}},\ \bibinfo
  {pages} {205127} (\bibinfo {year} {2014})}\BibitemShut {NoStop}%
\bibitem [{\citenamefont {Thakurathi}\ \emph {et~al.}(2013)\citenamefont
  {Thakurathi}, \citenamefont {Patel}, \citenamefont {Sen},\ and\ \citenamefont
  {Dutta}}]{FMaj3}%
  \BibitemOpen
  \bibfield  {author} {\bibinfo {author} {\bibfnamefont {M.}~\bibnamefont
  {Thakurathi}}, \bibinfo {author} {\bibfnamefont {A.~A.}\ \bibnamefont
  {Patel}}, \bibinfo {author} {\bibfnamefont {D.}~\bibnamefont {Sen}},\ and\
  \bibinfo {author} {\bibfnamefont {A.}~\bibnamefont {Dutta}},\ }\href
  {https://doi.org/10.1103/PhysRevB.88.155133} {\bibfield  {journal} {\bibinfo
  {journal} {Phys. Rev. B}\ }\textbf {\bibinfo {volume} {88}},\ \bibinfo
  {pages} {155133} (\bibinfo {year} {2013})}\BibitemShut {NoStop}%
\bibitem [{\citenamefont {Mondal}\ \emph {et~al.}(2023)\citenamefont {Mondal},
  \citenamefont {Ghosh}, \citenamefont {Nag},\ and\ \citenamefont
  {Saha}}]{FMaj4}%
  \BibitemOpen
  \bibfield  {author} {\bibinfo {author} {\bibfnamefont {D.}~\bibnamefont
  {Mondal}}, \bibinfo {author} {\bibfnamefont {A.~K.}\ \bibnamefont {Ghosh}},
  \bibinfo {author} {\bibfnamefont {T.}~\bibnamefont {Nag}},\ and\ \bibinfo
  {author} {\bibfnamefont {A.}~\bibnamefont {Saha}},\ }\href
  {https://doi.org/10.1103/PhysRevB.107.035427} {\bibfield  {journal} {\bibinfo
   {journal} {Phys. Rev. B}\ }\textbf {\bibinfo {volume} {107}},\ \bibinfo
  {pages} {035427} (\bibinfo {year} {2023})}\BibitemShut {NoStop}%
\bibitem [{\citenamefont {Liu}\ \emph {et~al.}(2013)\citenamefont {Liu},
  \citenamefont {Levchenko},\ and\ \citenamefont {Baranger}}]{FMaj5}%
  \BibitemOpen
  \bibfield  {author} {\bibinfo {author} {\bibfnamefont {D.~E.}\ \bibnamefont
  {Liu}}, \bibinfo {author} {\bibfnamefont {A.}~\bibnamefont {Levchenko}},\
  and\ \bibinfo {author} {\bibfnamefont {H.~U.}\ \bibnamefont {Baranger}},\
  }\href {https://doi.org/10.1103/PhysRevLett.111.047002} {\bibfield  {journal}
  {\bibinfo  {journal} {Phys. Rev. Lett.}\ }\textbf {\bibinfo {volume} {111}},\
  \bibinfo {pages} {047002} (\bibinfo {year} {2013})}\BibitemShut {NoStop}%
\bibitem [{\citenamefont {Thakurathi}\ \emph {et~al.}(2017)\citenamefont
  {Thakurathi}, \citenamefont {Loss},\ and\ \citenamefont {Klinovaja}}]{FMaj6}%
  \BibitemOpen
  \bibfield  {author} {\bibinfo {author} {\bibfnamefont {M.}~\bibnamefont
  {Thakurathi}}, \bibinfo {author} {\bibfnamefont {D.}~\bibnamefont {Loss}},\
  and\ \bibinfo {author} {\bibfnamefont {J.}~\bibnamefont {Klinovaja}},\ }\href
  {https://doi.org/10.1103/PhysRevB.95.155407} {\bibfield  {journal} {\bibinfo
  {journal} {Phys. Rev. B}\ }\textbf {\bibinfo {volume} {95}},\ \bibinfo
  {pages} {155407} (\bibinfo {year} {2017})}\BibitemShut {NoStop}%
\bibitem [{\citenamefont {Liu}\ \emph {et~al.}(2019)\citenamefont {Liu},
  \citenamefont {Shabani},\ and\ \citenamefont {Mitra}}]{FMaj7}%
  \BibitemOpen
  \bibfield  {author} {\bibinfo {author} {\bibfnamefont {D.~T.}\ \bibnamefont
  {Liu}}, \bibinfo {author} {\bibfnamefont {J.}~\bibnamefont {Shabani}},\ and\
  \bibinfo {author} {\bibfnamefont {A.}~\bibnamefont {Mitra}},\ }\href
  {https://doi.org/10.1103/PhysRevB.99.094303} {\bibfield  {journal} {\bibinfo
  {journal} {Phys. Rev. B}\ }\textbf {\bibinfo {volume} {99}},\ \bibinfo
  {pages} {094303} (\bibinfo {year} {2019})}\BibitemShut {NoStop}%
\bibitem [{\citenamefont {Yang}\ \emph {et~al.}(2021)\citenamefont {Yang},
  \citenamefont {Yang}, \citenamefont {Hu},\ and\ \citenamefont {Liu}}]{FMaj8}%
  \BibitemOpen
  \bibfield  {author} {\bibinfo {author} {\bibfnamefont {Z.}~\bibnamefont
  {Yang}}, \bibinfo {author} {\bibfnamefont {Q.}~\bibnamefont {Yang}}, \bibinfo
  {author} {\bibfnamefont {J.}~\bibnamefont {Hu}},\ and\ \bibinfo {author}
  {\bibfnamefont {D.~E.}\ \bibnamefont {Liu}},\ }\href
  {https://doi.org/10.1103/PhysRevLett.126.086801} {\bibfield  {journal}
  {\bibinfo  {journal} {Phys. Rev. Lett.}\ }\textbf {\bibinfo {volume} {126}},\
  \bibinfo {pages} {086801} (\bibinfo {year} {2021})}\BibitemShut {NoStop}%
\bibitem [{\citenamefont {Zhang}\ and\ \citenamefont
  {Das~Sarma}(2021)}]{FMaj9}%
  \BibitemOpen
  \bibfield  {author} {\bibinfo {author} {\bibfnamefont {R.-X.}\ \bibnamefont
  {Zhang}}\ and\ \bibinfo {author} {\bibfnamefont {S.}~\bibnamefont
  {Das~Sarma}},\ }\href {https://doi.org/10.1103/PhysRevLett.127.067001}
  {\bibfield  {journal} {\bibinfo  {journal} {Phys. Rev. Lett.}\ }\textbf
  {\bibinfo {volume} {127}},\ \bibinfo {pages} {067001} (\bibinfo {year}
  {2021})}\BibitemShut {NoStop}%
\bibitem [{\citenamefont {Kundu}\ and\ \citenamefont
  {Seradjeh}(2013)}]{FMaj10}%
  \BibitemOpen
  \bibfield  {author} {\bibinfo {author} {\bibfnamefont {A.}~\bibnamefont
  {Kundu}}\ and\ \bibinfo {author} {\bibfnamefont {B.}~\bibnamefont
  {Seradjeh}},\ }\href {https://doi.org/10.1103/PhysRevLett.111.136402}
  {\bibfield  {journal} {\bibinfo  {journal} {Phys. Rev. Lett.}\ }\textbf
  {\bibinfo {volume} {111}},\ \bibinfo {pages} {136402} (\bibinfo {year}
  {2013})}\BibitemShut {NoStop}%
\bibitem [{\citenamefont {Claassen}\ \emph {et~al.}(2019)\citenamefont
  {Claassen}, \citenamefont {Kennes}, \citenamefont {Zingl}, \citenamefont
  {Sentef},\ and\ \citenamefont {Rubio}}]{FTSC1}%
  \BibitemOpen
  \bibfield  {author} {\bibinfo {author} {\bibfnamefont {M.}~\bibnamefont
  {Claassen}}, \bibinfo {author} {\bibfnamefont {D.~M.}\ \bibnamefont
  {Kennes}}, \bibinfo {author} {\bibfnamefont {M.}~\bibnamefont {Zingl}},
  \bibinfo {author} {\bibfnamefont {M.~A.}\ \bibnamefont {Sentef}},\ and\
  \bibinfo {author} {\bibfnamefont {A.}~\bibnamefont {Rubio}},\ }\href
  {https://www.nature.com/articles/s41567-019-0532-6} {\bibfield  {journal}
  {\bibinfo  {journal} {Nat. Phys.}\ }\textbf {\bibinfo {volume} {15}},\
  \bibinfo {pages} {766} (\bibinfo {year} {2019})}\BibitemShut {NoStop}%
\bibitem [{\citenamefont {Kitamura}\ and\ \citenamefont {Aoki}(2022)}]{FTSC2}%
  \BibitemOpen
  \bibfield  {author} {\bibinfo {author} {\bibfnamefont {S.}~\bibnamefont
  {Kitamura}}\ and\ \bibinfo {author} {\bibfnamefont {H.}~\bibnamefont
  {Aoki}},\ }\href {https://www.nature.com/articles/s42005-022-00936-w}
  {\bibfield  {journal} {\bibinfo  {journal} {Commun. Phys.}\ }\textbf
  {\bibinfo {volume} {5}},\ \bibinfo {pages} {174} (\bibinfo {year}
  {2022})}\BibitemShut {NoStop}%
\bibitem [{\citenamefont {Haxell}\ \emph {et~al.}(2023)\citenamefont {Haxell},
  \citenamefont {Coraiola}, \citenamefont {Sabonis}, \citenamefont
  {Hinderling}, \citenamefont {Ten~Kate}, \citenamefont {Cheah}, \citenamefont
  {Krizek}, \citenamefont {Schott}, \citenamefont {Wegscheider}, \citenamefont
  {Belzig} \emph {et~al.}}]{FTSC3}%
  \BibitemOpen
  \bibfield  {author} {\bibinfo {author} {\bibfnamefont {D.~Z.}\ \bibnamefont
  {Haxell}}, \bibinfo {author} {\bibfnamefont {M.}~\bibnamefont {Coraiola}},
  \bibinfo {author} {\bibfnamefont {D.}~\bibnamefont {Sabonis}}, \bibinfo
  {author} {\bibfnamefont {M.}~\bibnamefont {Hinderling}}, \bibinfo {author}
  {\bibfnamefont {S.~C.}\ \bibnamefont {Ten~Kate}}, \bibinfo {author}
  {\bibfnamefont {E.}~\bibnamefont {Cheah}}, \bibinfo {author} {\bibfnamefont
  {F.}~\bibnamefont {Krizek}}, \bibinfo {author} {\bibfnamefont
  {R.}~\bibnamefont {Schott}}, \bibinfo {author} {\bibfnamefont
  {W.}~\bibnamefont {Wegscheider}}, \bibinfo {author} {\bibfnamefont
  {W.}~\bibnamefont {Belzig}}, \emph {et~al.},\ }\href
  {https://www.nature.com/articles/s41467-023-42357-5} {\bibfield  {journal}
  {\bibinfo  {journal} {Nat. Commun.}\ }\textbf {\bibinfo {volume} {14}},\
  \bibinfo {pages} {6798} (\bibinfo {year} {2023})}\BibitemShut {NoStop}%
\bibitem [{\citenamefont {Yao}\ \emph {et~al.}(2017)\citenamefont {Yao},
  \citenamefont {Yan},\ and\ \citenamefont {Wang}}]{FTSC4}%
  \BibitemOpen
  \bibfield  {author} {\bibinfo {author} {\bibfnamefont {S.}~\bibnamefont
  {Yao}}, \bibinfo {author} {\bibfnamefont {Z.}~\bibnamefont {Yan}},\ and\
  \bibinfo {author} {\bibfnamefont {Z.}~\bibnamefont {Wang}},\ }\href
  {https://doi.org/10.1103/PhysRevB.96.195303} {\bibfield  {journal} {\bibinfo
  {journal} {Phys. Rev. B}\ }\textbf {\bibinfo {volume} {96}},\ \bibinfo
  {pages} {195303} (\bibinfo {year} {2017})}\BibitemShut {NoStop}%
\bibitem [{\citenamefont {Morimoto}\ \emph {et~al.}(2017)\citenamefont
  {Morimoto}, \citenamefont {Po},\ and\ \citenamefont {Vishwanath}}]{FTSC5}%
  \BibitemOpen
  \bibfield  {author} {\bibinfo {author} {\bibfnamefont {T.}~\bibnamefont
  {Morimoto}}, \bibinfo {author} {\bibfnamefont {H.~C.}\ \bibnamefont {Po}},\
  and\ \bibinfo {author} {\bibfnamefont {A.}~\bibnamefont {Vishwanath}},\
  }\href {https://doi.org/10.1103/PhysRevB.95.195155} {\bibfield  {journal}
  {\bibinfo  {journal} {Phys. Rev. B}\ }\textbf {\bibinfo {volume} {95}},\
  \bibinfo {pages} {195155} (\bibinfo {year} {2017})}\BibitemShut {NoStop}%
\bibitem [{\citenamefont {Peng}\ and\ \citenamefont {Refael}(2019)}]{FTSC6}%
  \BibitemOpen
  \bibfield  {author} {\bibinfo {author} {\bibfnamefont {Y.}~\bibnamefont
  {Peng}}\ and\ \bibinfo {author} {\bibfnamefont {G.}~\bibnamefont {Refael}},\
  }\href {https://doi.org/10.1103/PhysRevLett.123.016806} {\bibfield  {journal}
  {\bibinfo  {journal} {Phys. Rev. Lett.}\ }\textbf {\bibinfo {volume} {123}},\
  \bibinfo {pages} {016806} (\bibinfo {year} {2019})}\BibitemShut {NoStop}%
\bibitem [{\citenamefont {Bazarnik}\ \emph {et~al.}(2023)\citenamefont
  {Bazarnik}, \citenamefont {Lo~Conte}, \citenamefont {Mascot}, \citenamefont
  {von Bergmann}, \citenamefont {Morr},\ and\ \citenamefont
  {Wiesendanger}}]{TNPSC}%
  \BibitemOpen
  \bibfield  {author} {\bibinfo {author} {\bibfnamefont {M.}~\bibnamefont
  {Bazarnik}}, \bibinfo {author} {\bibfnamefont {R.}~\bibnamefont {Lo~Conte}},
  \bibinfo {author} {\bibfnamefont {E.}~\bibnamefont {Mascot}}, \bibinfo
  {author} {\bibfnamefont {K.}~\bibnamefont {von Bergmann}}, \bibinfo {author}
  {\bibfnamefont {D.~K.}\ \bibnamefont {Morr}},\ and\ \bibinfo {author}
  {\bibfnamefont {R.}~\bibnamefont {Wiesendanger}},\ }\href
  {https://www.nature.com/articles/s41467-023-36201-z} {\bibfield  {journal}
  {\bibinfo  {journal} {Nat. Commun.}\ }\textbf {\bibinfo {volume} {14}},\
  \bibinfo {pages} {614} (\bibinfo {year} {2023})}\BibitemShut {NoStop}%
\bibitem [{\citenamefont {Li}\ \emph {et~al.}(2016)\citenamefont {Li},
  \citenamefont {Neupert}, \citenamefont {Wang}, \citenamefont {MacDonald},
  \citenamefont {Yazdani},\ and\ \citenamefont {Bernevig}}]{MHS1}%
  \BibitemOpen
  \bibfield  {author} {\bibinfo {author} {\bibfnamefont {J.}~\bibnamefont
  {Li}}, \bibinfo {author} {\bibfnamefont {T.}~\bibnamefont {Neupert}},
  \bibinfo {author} {\bibfnamefont {Z.}~\bibnamefont {Wang}}, \bibinfo {author}
  {\bibfnamefont {A.}~\bibnamefont {MacDonald}}, \bibinfo {author}
  {\bibfnamefont {A.}~\bibnamefont {Yazdani}},\ and\ \bibinfo {author}
  {\bibfnamefont {B.~A.}\ \bibnamefont {Bernevig}},\ }\href
  {https://www.nature.com/articles/ncomms12297} {\bibfield  {journal} {\bibinfo
   {journal} {Nat. Commun.}\ }\textbf {\bibinfo {volume} {7}},\ \bibinfo
  {pages} {12297} (\bibinfo {year} {2016})}\BibitemShut {NoStop}%
\bibitem [{\citenamefont {M{\'e}nard}\ \emph {et~al.}(2017)\citenamefont
  {M{\'e}nard}, \citenamefont {Guissart}, \citenamefont {Brun}, \citenamefont
  {Leriche}, \citenamefont {Trif}, \citenamefont {Debontridder}, \citenamefont
  {Demaille}, \citenamefont {Roditchev}, \citenamefont {Simon},\ and\
  \citenamefont {Cren}}]{MHS2}%
  \BibitemOpen
  \bibfield  {author} {\bibinfo {author} {\bibfnamefont {G.~C.}\ \bibnamefont
  {M{\'e}nard}}, \bibinfo {author} {\bibfnamefont {S.}~\bibnamefont
  {Guissart}}, \bibinfo {author} {\bibfnamefont {C.}~\bibnamefont {Brun}},
  \bibinfo {author} {\bibfnamefont {R.~T.}\ \bibnamefont {Leriche}}, \bibinfo
  {author} {\bibfnamefont {M.}~\bibnamefont {Trif}}, \bibinfo {author}
  {\bibfnamefont {F.}~\bibnamefont {Debontridder}}, \bibinfo {author}
  {\bibfnamefont {D.}~\bibnamefont {Demaille}}, \bibinfo {author}
  {\bibfnamefont {D.}~\bibnamefont {Roditchev}}, \bibinfo {author}
  {\bibfnamefont {P.}~\bibnamefont {Simon}},\ and\ \bibinfo {author}
  {\bibfnamefont {T.}~\bibnamefont {Cren}},\ }\href
  {https://www.nature.com/articles/s41467-017-02192-x} {\bibfield  {journal}
  {\bibinfo  {journal} {Nat. Commun.}\ }\textbf {\bibinfo {volume} {8}},\
  \bibinfo {pages} {2040} (\bibinfo {year} {2017})}\BibitemShut {NoStop}%
\bibitem [{\citenamefont {Kim}\ \emph {et~al.}(2018)\citenamefont {Kim},
  \citenamefont {Palacio-Morales}, \citenamefont {Posske}, \citenamefont
  {R{\'o}zsa}, \citenamefont {Palot{\'a}s}, \citenamefont {Szunyogh},
  \citenamefont {Thorwart},\ and\ \citenamefont {Wiesendanger}}]{MHS3}%
  \BibitemOpen
  \bibfield  {author} {\bibinfo {author} {\bibfnamefont {H.}~\bibnamefont
  {Kim}}, \bibinfo {author} {\bibfnamefont {A.}~\bibnamefont
  {Palacio-Morales}}, \bibinfo {author} {\bibfnamefont {T.}~\bibnamefont
  {Posske}}, \bibinfo {author} {\bibfnamefont {L.}~\bibnamefont {R{\'o}zsa}},
  \bibinfo {author} {\bibfnamefont {K.}~\bibnamefont {Palot{\'a}s}}, \bibinfo
  {author} {\bibfnamefont {L.}~\bibnamefont {Szunyogh}}, \bibinfo {author}
  {\bibfnamefont {M.}~\bibnamefont {Thorwart}},\ and\ \bibinfo {author}
  {\bibfnamefont {R.}~\bibnamefont {Wiesendanger}},\ }\href
  {https://www.science.org/doi/10.1126/sciadv.aar5251} {\bibfield  {journal}
  {\bibinfo  {journal} {Sci. Adv.}\ }\textbf {\bibinfo {volume} {4}},\ \bibinfo
  {pages} {eaar5251} (\bibinfo {year} {2018})}\BibitemShut {NoStop}%
\bibitem [{\citenamefont {Palacio-Morales}\ \emph {et~al.}(2019)\citenamefont
  {Palacio-Morales}, \citenamefont {Mascot}, \citenamefont {Cocklin},
  \citenamefont {Kim}, \citenamefont {Rachel}, \citenamefont {Morr},\ and\
  \citenamefont {Wiesendanger}}]{MHS4}%
  \BibitemOpen
  \bibfield  {author} {\bibinfo {author} {\bibfnamefont {A.}~\bibnamefont
  {Palacio-Morales}}, \bibinfo {author} {\bibfnamefont {E.}~\bibnamefont
  {Mascot}}, \bibinfo {author} {\bibfnamefont {S.}~\bibnamefont {Cocklin}},
  \bibinfo {author} {\bibfnamefont {H.}~\bibnamefont {Kim}}, \bibinfo {author}
  {\bibfnamefont {S.}~\bibnamefont {Rachel}}, \bibinfo {author} {\bibfnamefont
  {D.~K.}\ \bibnamefont {Morr}},\ and\ \bibinfo {author} {\bibfnamefont
  {R.}~\bibnamefont {Wiesendanger}},\ }\href
  {https://www.science.org/doi/10.1126/sciadv.aav6600} {\bibfield  {journal}
  {\bibinfo  {journal} {Sci. Adv.}\ }\textbf {\bibinfo {volume} {5}},\ \bibinfo
  {pages} {eaav6600} (\bibinfo {year} {2019})}\BibitemShut {NoStop}%
\bibitem [{\citenamefont {Kezilebieke}\ \emph {et~al.}(2020)\citenamefont
  {Kezilebieke}, \citenamefont {Huda}, \citenamefont {Va{\v{n}}o},
  \citenamefont {Aapro}, \citenamefont {Ganguli}, \citenamefont {Silveira},
  \citenamefont {G{\l}odzik}, \citenamefont {Foster}, \citenamefont {Ojanen},\
  and\ \citenamefont {Liljeroth}}]{MHS5}%
  \BibitemOpen
  \bibfield  {author} {\bibinfo {author} {\bibfnamefont {S.}~\bibnamefont
  {Kezilebieke}}, \bibinfo {author} {\bibfnamefont {M.~N.}\ \bibnamefont
  {Huda}}, \bibinfo {author} {\bibfnamefont {V.}~\bibnamefont {Va{\v{n}}o}},
  \bibinfo {author} {\bibfnamefont {M.}~\bibnamefont {Aapro}}, \bibinfo
  {author} {\bibfnamefont {S.~C.}\ \bibnamefont {Ganguli}}, \bibinfo {author}
  {\bibfnamefont {O.~J.}\ \bibnamefont {Silveira}}, \bibinfo {author}
  {\bibfnamefont {S.}~\bibnamefont {G{\l}odzik}}, \bibinfo {author}
  {\bibfnamefont {A.~S.}\ \bibnamefont {Foster}}, \bibinfo {author}
  {\bibfnamefont {T.}~\bibnamefont {Ojanen}},\ and\ \bibinfo {author}
  {\bibfnamefont {P.}~\bibnamefont {Liljeroth}},\ }\href
  {https://www.nature.com/articles/s41586-020-2989-y} {\bibfield  {journal}
  {\bibinfo  {journal} {Nature}\ }\textbf {\bibinfo {volume} {588}},\ \bibinfo
  {pages} {424} (\bibinfo {year} {2020})}\BibitemShut {NoStop}%
\bibitem [{\citenamefont {Mascot}\ \emph {et~al.}(2021)\citenamefont {Mascot},
  \citenamefont {Bedow}, \citenamefont {Graham}, \citenamefont {Rachel},\ and\
  \citenamefont {Morr}}]{MHS6}%
  \BibitemOpen
  \bibfield  {author} {\bibinfo {author} {\bibfnamefont {E.}~\bibnamefont
  {Mascot}}, \bibinfo {author} {\bibfnamefont {J.}~\bibnamefont {Bedow}},
  \bibinfo {author} {\bibfnamefont {M.}~\bibnamefont {Graham}}, \bibinfo
  {author} {\bibfnamefont {S.}~\bibnamefont {Rachel}},\ and\ \bibinfo {author}
  {\bibfnamefont {D.~K.}\ \bibnamefont {Morr}},\ }\href
  {https://www.nature.com/articles/s41535-020-00299-x} {\bibfield  {journal}
  {\bibinfo  {journal} {npj Quantum Mater.}\ }\textbf {\bibinfo {volume} {6}},\
  \bibinfo {pages} {6} (\bibinfo {year} {2021})}\BibitemShut {NoStop}%
\bibitem [{\citenamefont {Wong}\ \emph {et~al.}(2023)\citenamefont {Wong},
  \citenamefont {Hirsbrunner}, \citenamefont {Gliozzi}, \citenamefont {Malik},
  \citenamefont {Bradlyn}, \citenamefont {Hughes},\ and\ \citenamefont
  {Morr}}]{MHS7}%
  \BibitemOpen
  \bibfield  {author} {\bibinfo {author} {\bibfnamefont {K.~H.}\ \bibnamefont
  {Wong}}, \bibinfo {author} {\bibfnamefont {M.~R.}\ \bibnamefont
  {Hirsbrunner}}, \bibinfo {author} {\bibfnamefont {J.}~\bibnamefont
  {Gliozzi}}, \bibinfo {author} {\bibfnamefont {A.}~\bibnamefont {Malik}},
  \bibinfo {author} {\bibfnamefont {B.}~\bibnamefont {Bradlyn}}, \bibinfo
  {author} {\bibfnamefont {T.~L.}\ \bibnamefont {Hughes}},\ and\ \bibinfo
  {author} {\bibfnamefont {D.~K.}\ \bibnamefont {Morr}},\ }\href
  {https://www.nature.com/articles/s41535-023-00564-9} {\bibfield  {journal}
  {\bibinfo  {journal} {npj Quantum Mater.}\ }\textbf {\bibinfo {volume} {8}},\
  \bibinfo {pages} {31} (\bibinfo {year} {2023})}\BibitemShut {NoStop}%
\bibitem [{\citenamefont {Kieu}\ \emph {et~al.}(2023)\citenamefont {Kieu},
  \citenamefont {Mascot}, \citenamefont {Bedow}, \citenamefont {Wiesendanger},\
  and\ \citenamefont {Morr}}]{MHS8}%
  \BibitemOpen
  \bibfield  {author} {\bibinfo {author} {\bibfnamefont {T.}~\bibnamefont
  {Kieu}}, \bibinfo {author} {\bibfnamefont {E.}~\bibnamefont {Mascot}},
  \bibinfo {author} {\bibfnamefont {J.}~\bibnamefont {Bedow}}, \bibinfo
  {author} {\bibfnamefont {R.}~\bibnamefont {Wiesendanger}},\ and\ \bibinfo
  {author} {\bibfnamefont {D.~K.}\ \bibnamefont {Morr}},\ }\href
  {https://doi.org/10.1103/PhysRevB.108.L060509} {\bibfield  {journal}
  {\bibinfo  {journal} {Phys. Rev. B}\ }\textbf {\bibinfo {volume} {108}},\
  \bibinfo {pages} {L060509} (\bibinfo {year} {2023})}\BibitemShut {NoStop}%
\bibitem [{\citenamefont {Sancho}\ \emph {et~al.}(1985)\citenamefont {Sancho},
  \citenamefont {Sancho}, \citenamefont {Sancho},\ and\ \citenamefont
  {Rubio}}]{RGFM}%
  \BibitemOpen
  \bibfield  {author} {\bibinfo {author} {\bibfnamefont {M.~L.}\ \bibnamefont
  {Sancho}}, \bibinfo {author} {\bibfnamefont {J.~L.}\ \bibnamefont {Sancho}},
  \bibinfo {author} {\bibfnamefont {J.~L.}\ \bibnamefont {Sancho}},\ and\
  \bibinfo {author} {\bibfnamefont {J.}~\bibnamefont {Rubio}},\ }\href
  {https://doi.org/10.1088/0305-4608/15/4/009} {\bibfield  {journal} {\bibinfo
  {journal} {J. Phys. F: Met. Phys.}\ }\textbf {\bibinfo {volume} {15}},\
  \bibinfo {pages} {851} (\bibinfo {year} {1985})}\BibitemShut {NoStop}%
\end{thebibliography}

%

\end{document}